\documentclass[useAMS,usenatbib,asm]{aa}

\usepackage[varg]{txfonts}
\usepackage[english]{babel}
\usepackage{graphicx}   
\usepackage[space]{grffile}     
\usepackage{amsmath}
\usepackage{amssymb}
\usepackage{color}

\begin{document}


\title{Warp evidences in precessing galactic bar models}

\author{P. S\'anchez-Mart\'{i}n\inst{\ref{inst1},\ref{inst3}}\and M. Romero-G\'{o}mez\inst{\ref{inst2}}\and
J. J. Masdemont\inst{\ref{inst1}}}

\institute{IEEC-UPC i Dept. de Matem\`{a}tiques, Universitat Polit\`{e}cnica de 
Catalunya, Diagonal 647 (ETSEIB), E08028 Barcelona, Spain, \email{patricia.sanchez.martin@upc.edu, josep.masdemont@upc.edu}\label{inst1} \and Dept. d'Astronomia i Meteorologia, Institut de Ci\`{e}ncies del Cosmos, Universitat de Barcelona, IEEC, Mart\'{i} i Franqu\`{e}s 1, E08028 Barcelona, Spain, \email{mromero{@}am.ub.es}\label{inst2} \and GTM - Grup de recerca en Tecnologies M\`edia, La Salle, Universitat Ramon Llull, Quatre Camins 2, E08022 Barcelona, Spain, \email{psanchez@salleurl.edu}\label{inst3}
} 

\date{}

\abstract{
Most galaxies have a warped shape when they are seen from an edge-on point of
view. The reason for this curious form is not completely known so far and in
this work we apply dynamical system tools to contribute to its explanation.
Starting from a simple, but realistic, model formed by a bar and a disc, we
study the effect produced by a small misalignment between the angular momentum
of the system and its angular velocity. To this end, a precession model is
developed and considered, assuming that the bar behaves like a rigid body.
After checking that the periodic orbits inside the bar keep being the skeleton
of the inner system, even after inflicting a precession to the potential, we
compute the invariant manifolds of the unstable periodic orbits departing from
the equilibrium points at the ends of the bar to get evidences of their warped
shapes. As it is well known, the invariant manifolds associated with these
periodic orbits drive the arms and rings of barred galaxies and constitute the
skeleton of these building blocks. Looking at them from a side-on viewpoint,
we find that these manifolds present warped shapes as those recognized in
observations. Lastly, test particle simulations have been performed to determine
how the stars are affected by the applied precession, confirming this way the
theoretical results obtained.
}


\maketitle


\section{Introduction}

\begin{figure}
  \centering
    \includegraphics[width=0.4\textwidth]{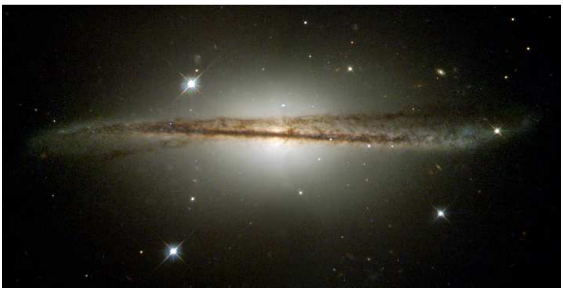}
  \caption{Galaxy ESO 510-G13 photographed by Hubble telescope.}
  \label{fig:warp_galaxia}
\end{figure}

In this study we focus on the warps observed in some galaxies when seen 
edge-on.
Thanks to the images of the Hubble Space Telescope
and taking into account the probability of non-detection
of warps when the line of nodes lies in the plane of the sky, 
it has been observed that nearly all galaxies are warped, confirming the 
suggestion made by \citet{Bosma} for HI warps~\citep{catalog}.
Although there is abundant literature about this subject, the reason for these
warps is not known yet. They have been observed in the distribution of stars
\citep[e.g.][]{Sanchez-Saavedra} and in the study of neutral
hydrogen~\citep{Bosma}, confirming that they are a very common phenomenon. In
general, warps are commonly viewed as an integral sign as seen edge-on,
manifesting themselves in the shape of the outer disc and bending away from the
plane defined by the inner disc (like the galaxy shown in
Fig.~\ref{fig:warp_galaxia}). This fact suggested some misalignment between
the angular momenta of
some material in warps and some material in the inner disc.
In this direction, \citet{DebSell} made simulations where the warp was formed
when a misalignment between the angular momenta of the disc and the halo
occurs.

Several assumptions have been made in the literature about the formation of
warps. \citet{Briggs} established some rules based on observational studies of
external galaxies to determine the behaviour of galactic warps and claimed that
warps appear from isophotal radius $R_{26.5}$. After some time, \citet{Cox}
studied the observations of the galaxy UGC 7170 concluding that due to the
similarities between the stellar and gaseous warps, these could be produced by
means of a gravitational origin. And more recently, \citet{Sellwood} determined
that since warps are really common, they should be either repeatedly
regenerated or long-lived.

From a theoretical point of view, numerous approaches have been made to
understand the mechanisms responsible for warp generation. One of these
mechanisms, explained in \citet{Lynden-Bell}, established that warps could be
produced by internal bending modes in the disc as a long-lived phenomenon, but
this proposal held only for a disc with an unrealistic mass truncation
\citep{HunterToomre}. In this context, \citet{SparkeCasertano} found warp modes
inside rigid halos by means of discrete modes of bending. In \citet{bend_warp}
this discussion was revived and identified short-lived bending instabilities as
a possible cause of the formation of warps, obtaining however, warp angles of
less than 5$^{\circ}$. Nevertheless, \citet{Binney98} argued that the inner
halo would realign with the disc and so the warp would dissipate. 

Another possible explanation for the existence of warps is a tidal interaction
between galaxies, this theory has been studied to explain the Milky Way's warp,
mainly because of the proximity of the Large Magellanic Clouds \citep[see
e.g.][]{AvnerKing, HunterToomre, Levine}. Nevertheless,
\citet{LopezCorredoira} developed a method to calculate the amplitude of the 
galactic
warp generated by a torque due to external forces. The method was applied to
discard the tidal theory since it would lead to the formation of
warps of very low amplitude. In fact, the authors proposed that  warps are 
formed due to the accretion of material over the disc (i.e. the accretion of
angular momentum). \citet{Read} agreed that the warp is an indicator of the
merger activity and stated that if the impact angle were larger than $20^\circ$ 
then the stellar disc could be warped. 

As we can see, the explanation of the existence of warped galaxies 
represents a challenge. Our approach is based on the fact that warps are
long-lived. By means of dynamical system tools, we take a simple but widely
used model, composed by a bar and a disc, to show that introducing a natural
misalignment between the angular momentum and the angular velocity, the model
is consistent and it is able to reproduce warped shapes. To check its
consistency, we apply the fact that the periodic orbits are the backbone of
galactic bars since these orbits are mainly stable and therefore they mostly
determine the structure of the bar \citep{ContopPapayan, Athan1983,
ContopoulosGrosbol}. 

Then we study the set of orbits that depart from the Lyapunov orbits, proving
that they acquire the warped shape with varying angles. The line of our study
follows the analysis of the invariant objects which, in a similar dynamic way,
cause the formation of rings and spiral arms in barred galaxies
\citep{VoglisStavr, Voglis2006, Patsis, Romero1,Romero2,Romero3}. Again the
purpose is to study the invariant manifolds associated to these objects. How
these manifolds are affected by a precessing model and they explain this way
the appearance of galactic warps. Since invariant manifolds are determined by
the potential of the galaxy, they exist for as long as this potential does not
change significantly and thus they are long-lived objects.

In Section~\ref{sect:section2} we justify the consistency of the misalignment
between the angular momentum and angular velocity vectors and we derive the
equations of motion of the precessing model. We also describe the galactic
potential considered and the characteristics of this potential in the
precessing model. In Section~\ref{sect:section3} we detail the types of orbits
inside the bar, we observe how they are modified with the tilt angle and we
check that they are the skeleton of the model. The formation of warps in
our model is described in Section~\ref{sect:section4} where we study the
invariant manifolds of the system that, as it is well known, provide the
backbone of dynamics.  In Section~\ref{sect:section5} we perform test-particle
simulations of the model which, although they are collisionless simulations,
they serve to demonstrate that the orbits in the model behave as predicted.  
Finally, conclusions and expectations for future work are given in
Section~\ref{sect:section6}.

\section{The precessing model} \label{sect:section2}

There are some theories which seek to explain the formation of warps through a
misalignment between the angular momenta of the components of the models.
Usually they assume that this is produced by the contribution of a third
element. For instance, as an accretion of material due to the cosmic
infall~\citep[see e.g.][]{Ostriker, Jiang, LopezCorredoira}, and also possibly
aided by dynamical friction between the 
components~\citep[see e.g.][]{DebSell}.  Although this could be one of the 
reasons, the dynamics in the formation of the bar and other blocks of the
galaxy could lead to a small misalignment between the angular velocity,
$\boldsymbol{\omega}$, and the angular momentum, $\mathbf{L}$, without any additional
perturbation. 

If we understand the galaxy and the formation of the bar as an accretion of
material from a spinning mass distribution, the total angular momentum will be
preserved during the process. However, for the angular velocity of its building
blocks, even though the main component will be in the direction of the angular
momentum, a small component can appear in the orthogonal direction causing the
misalignment. Probabilistically speaking, it is natural that it occurs in this
way,  moreover it is reinforced by the existence of any other external
perturbations or internal frictions. This is, the result of having angular
momentum and angular velocity slightly misaligned in the motion of the bar
should be a common phenomenon even when considering torque free motions of rigid
bodies. In fact, the probability that $\mathbf{L}$ and $\boldsymbol{\omega}$ are
aligned is very small, if not zero. The result of this misalignment is a small
precession of the bar.

\citet{Combes} hypothesized about the relation between the precession of the
angular momentum of a galactic disc and the formation of a warp as a vibration
of the disc, but she did not pursue this idea any further. Although we reached 
the idea of studying the misalignment of the angular momentum and the angular
velocity independently of \citet{Combes}, the work carried out in this paper
may be considered, to some extent, a detailed study of the original question.
Considering a galactic model formed by a bar and a disc, the main purpose of
this paper is to study the effect of a small misalignment between the angular 
momentum of the system and its angular velocity. 

The fundamentals of the motion of rigid bodies can be found in many books
of classical mechanics~\citep[a classical reference e.g.][]{Goldstein}, 
in what follows we summarize just a few main concepts that we need in order
to introduce a precessing bar in a usual galactic model.

The main equation in rigid body dynamics relating angular momentum and
angular velocity when they are expressed in a reference frame attached
to the body is $\mathbf{L}=\mathbf{I}\cdot\boldsymbol{\omega}$, where 
$\mathbf{I}$ is known as the inertia tensor. For the considered body frame,
the tensor $\mathbf{I}$ is a symmetric constant matrix whose values depend only
on the mass distribution. Since it is symmetric, this inertia tensor can be
written in diagonal form,
$\mathbf{I}_b=\hbox{\rm diag}\{I_1,I_2,I_3\}$, in an orthogonal basis. Its
eigenvectors point to what are known as the principal axes of rotation while the
components $I_1$, $I_2$ and $I_3$ are called principal moments of inertia.  In
case of bodies with symmetric mass distributions, the principal axes are
related with the symmetries. For instance, in case of a Ferrers bar \citep{Ferrers},
independently of its degree of homogeneity, our first principal axis is aligned
with the major axis of the bar in the $x$ direction while the remaining two
ones are aligned with the major and minor axis of the ellipse obtained when
cutting the Ferrers ellipsoid with the plane $x=0$.  Moreover, even though we
will not be restricted to this case, for a constant density ellipsoid, the
principal moments of inertia are given by, $I_1 = \frac{1}{5}M_b(b^2+c^2)$,
$I_2=\frac{1}{5}M_b(a^2+c^2)$ and $I_3=\frac{1}{5}M_b(a^2+b^2)$, where $M_b$
denotes the mass of the bar, and the parameters $a$ (semi-major axis) and $b$,
$c$ (intermediate and semi-minor axes, respectively) define the shape of the bar.  

\begin{figure}
  \centering
    \includegraphics[width=0.35\textwidth]{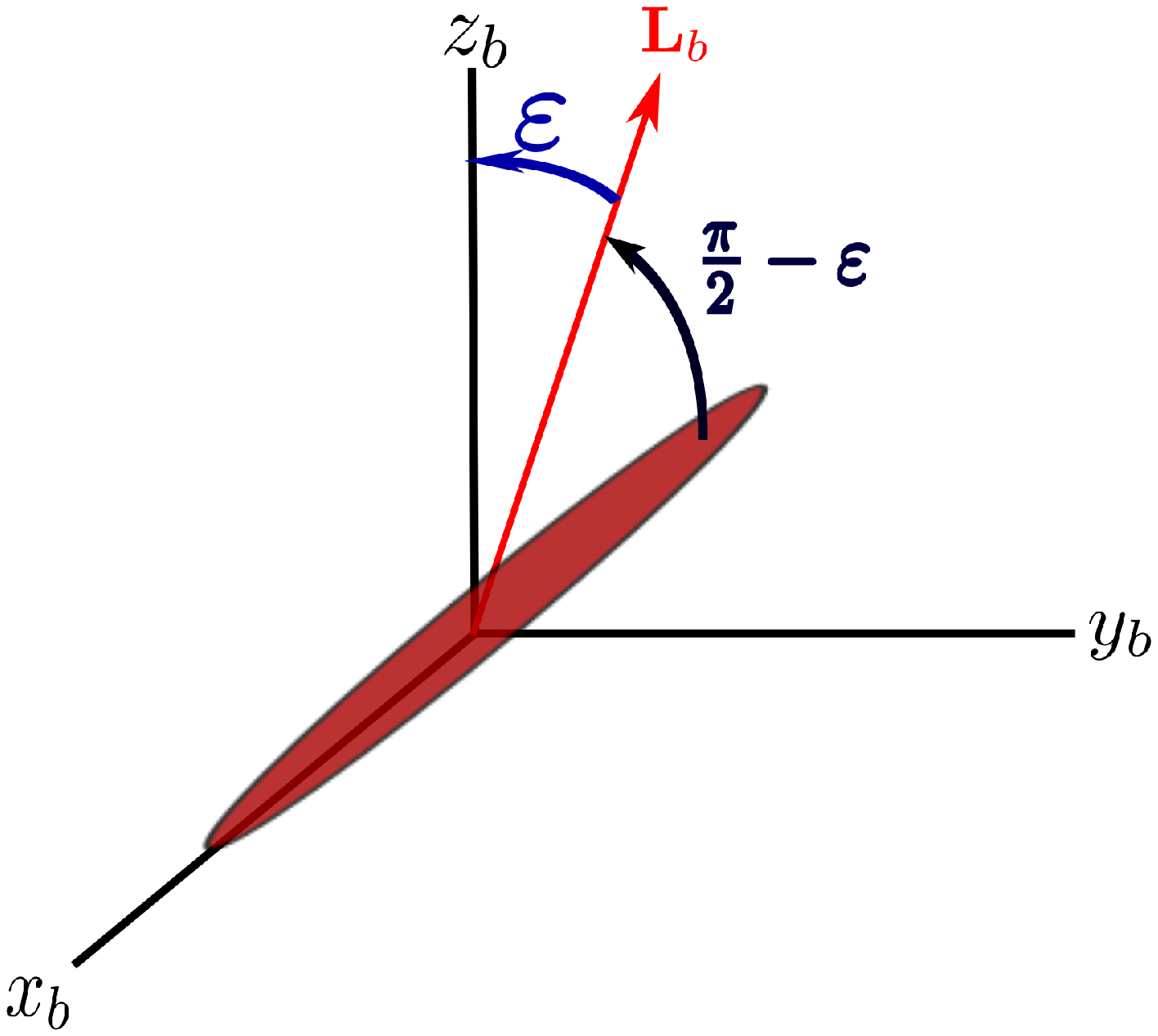}
    \includegraphics[width=0.35\textwidth]{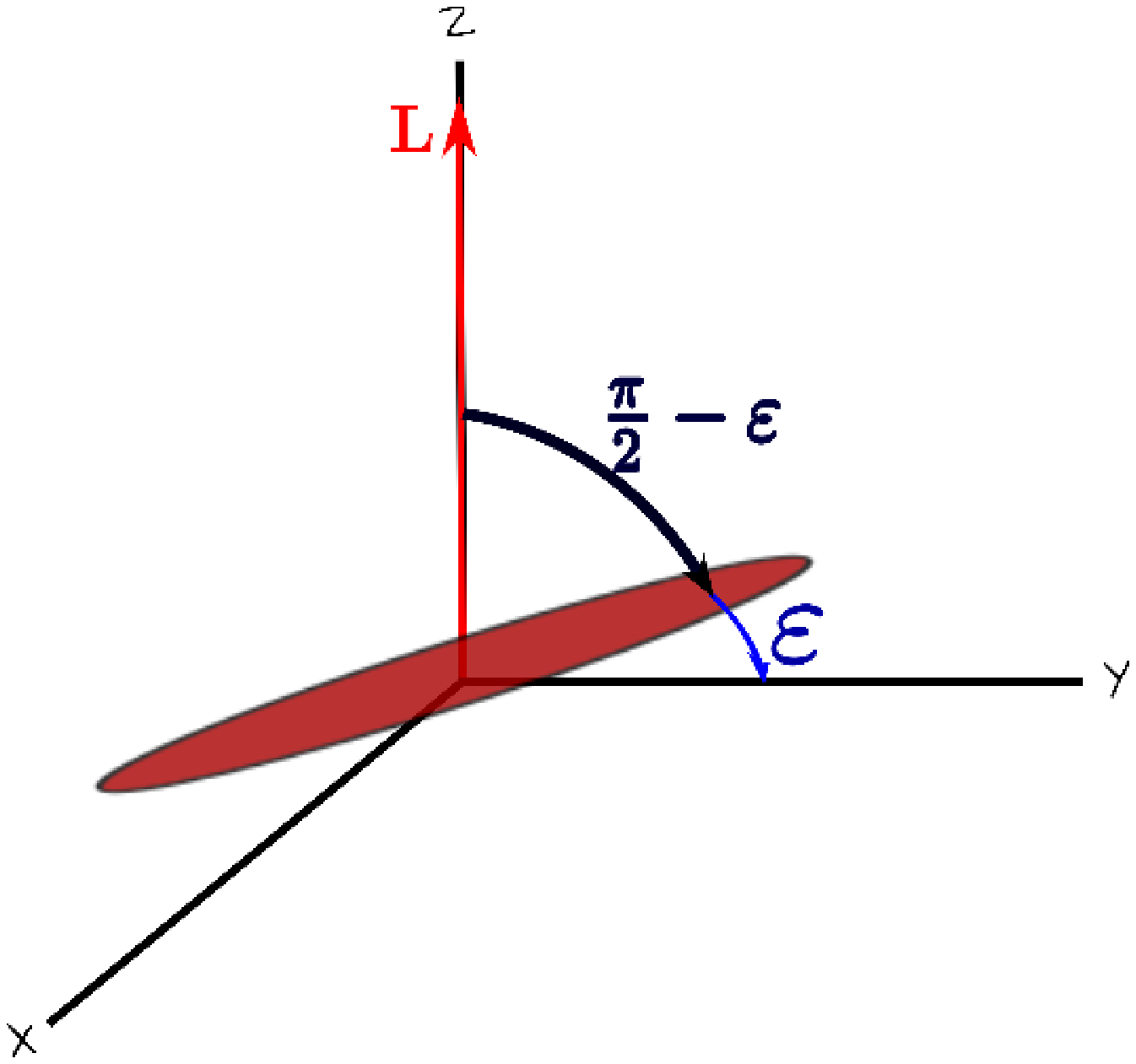}
    \caption{Reference systems. Top: Bar in the body reference frame, where the angular momentum $\mathbf{L}_b$ and the $y_bz_b$ plane keep an angle $\varepsilon$. Bottom: Bar in the inertial reference frame, where the angular momentum is aligned with the $Z$ axis and the major axis of the bar keeps an angle $\varepsilon$ with the $XY$ plane.}
      \label{fig:bar_bodyandiner}
\end{figure}

Another main ingredient to model the motion of our precessing bar are the
Euler equations. Assuming a torque-free motion, 
the angular velocity of the bar with respect to the static inertial axes,
but expressed in the body frame (whose axes are aligned with the principal axes 
of the bar, see top panel of Fig.~\ref{fig:bar_bodyandiner}), $\boldsymbol{\omega}_b=(\omega_1,\omega_2,\omega_3)$, is a solution of Euler's
equations:
\begin{equation}
\left\lbrace
\begin{array}{l}
 I_1\frac{d\omega_1}{dt} = \omega_2 \omega_3 (I_2 - I_3), \\
 I_2\frac{d\omega_2}{dt} = \omega_1 \omega_3 (I_3 - I_1), \\
 I_3\frac{d\omega_3}{dt} = \omega_1 \omega_2 (I_1 - I_2). \\
 \end{array}
\right.
 \label{eqn:euler}
\end{equation}

We are going to study Eq.~\eqref{eqn:euler} in the case of an axially symmetric
bar along the $x$ axis, with parameters $a>b=c$. Since the major axis of the
bar is along the $x$ axis in body coordinates, we have $I_1 \neq I_2 = I_3$ 
and immediately it follows that $\omega_1$ and $A^2=\omega_2^2+\omega_3^2$ are 
constants of the motion. Then the angular velocity of the bar expressed in the 
body frame is,
\begin{equation}
  \boldsymbol{\omega}_b = \left(
			    \begin{array}{l}
			      \omega_1 \\
			       A \sin(\lambda t) \\
			       A \cos(\lambda t) \\
			    \end{array}
			    \right).
\label{eqn:LenVelang}
\end{equation}
where we have defined,
\begin{equation}
 \lambda := \frac{I_T-I_1}{I_T}\omega_1, \qquad (I_T := I_2 = I_3).
 \label{eqn:lambda}
\end{equation}
Note that $\lambda$ is the precession rate of $\boldsymbol{\omega}_b$ in a cone 
around the main axis of the bar.

The angular momentum expressed in the body frame, 
$\mathbf{L}_b = \mathbf{I}_b\cdot\boldsymbol{\omega}_b$, has constant modulus
$L=||\mathbf{L}_b||=\sqrt{I_1^2\omega_1^2+I_T^2A^2}$ and describes a cone
about the $x$ axis. Let $\varepsilon$ be the angle from $\mathbf{L}_b$ to the 
$yz$ plane in the body reference (this is, the angle between the generatrix of
the cone, $\mathbf{L}_b$, and the negative $x$ semi-axis is $\frac{\pi}{2}-\varepsilon$,
as is represented in the top panel of Fig.~\ref{fig:bar_bodyandiner}).
Let us remark that in our study $\varepsilon$ will be always a small parameter. 
In terms of $L$ and $\varepsilon$, the angular momentum of the bar in the body 
frame can be written as,
\begin{equation}
  \mathbf{L}_b = \left(
		    \begin{array}{l}
		      -L\sin(\varepsilon) \\
		      L\cos(\varepsilon) \sin(\lambda t) \\
		      L\cos(\varepsilon) \cos(\lambda t) \\
		    \end{array}
		  \right).
\label{eqn:LenMomangfin}
\end{equation}
Again we note that $\mathbf{L}_b$ has a small constant component,
$-L\sin(\varepsilon)$, in the $x$ direction of the body frame and a big one of
modulus $L\cos(\varepsilon)$ describing a circle in the $yz$ plane of the body
frame. The angular velocity in the body frame is given by,
\begin{equation}
  \boldsymbol{\omega}_b = \mathbf{I}_b^{-1}\cdot\mathbf{L}_b = \left(
			    \begin{array}{l}
			      -\frac{L}{I_1}\sin(\varepsilon) \\
			      \frac{L}{I_T}\cos(\varepsilon) \sin(\lambda t) \\
			      \frac{L}{I_T}\cos(\varepsilon)\cos(\lambda t) \\
			    \end{array}
			    \right).
\label{eqn:LenOmfin}
\end{equation}
It follows from Eqs.~\eqref{eqn:lambda} and \eqref{eqn:LenOmfin} that $\lambda
= -\frac{I_T-I_1}{I_T}\frac{L}{I_1}\sin(\varepsilon)$. Since the main axis,
$a$, of the ellipsoid is much greater than the other axes, $b$ and $c$, we have
that $I_1 < I_T$. Moreover, for small values of $\varepsilon$,
$\sin(\varepsilon) \approx \varepsilon$, and therefore $\lambda$ is also small. 
Finally let us note that when $\varepsilon =0$, then $\lambda =0$ and both, 
angular momentum and the angular velocity are aligned on the $z$ axis.

Since in an inertial reference frame the angular momentum, $\mathbf{L}$,
is preserved, let us take the inertial system so that the $Z$ axis is 
aligned with $\mathbf{L}$. The major axis of the bar has to keep an angle
$\frac{\pi}{2}-\varepsilon$ with respect $\mathbf{L}$ and then $\varepsilon$ 
also measures the angle between the main axis of the bar and the $XY$ plane 
of the inertial reference system.  For this reason, from now on, it will be 
referred as the \emph{tilt angle} of the motion of the bar (see bottom panel of
Fig.~\ref{fig:bar_bodyandiner}). Note that in the inertial frame the major
axis of the bar describes a cone about the $Z$ axis, while $\mathbf{L}$ and $\boldsymbol{\omega}$ 
are slightly misaligned since the bar is also rotating about its major axis.
When $\varepsilon =0$ the major axis of the bar rotates inside the $XY$ plane,
there is no rotation of the bar about its major axis and again, $\mathbf{L}$ and $\boldsymbol{\omega}$ are 
aligned.


We define now a new non-inertial reference frame henceforth called the
\emph{precessing reference system}. In this reference frame the $x$ axis is
aligned with the major axis of the bar, but the bar is not fixed like in
the body frame but rotating around the $x$ axis.

Note that, in the body reference system, the angular momentum and angular
velocity vectors rotate around the main axis of the bar ($x$ axis in the body
frame) with angular speed $\lambda$. The precessing reference frame 
rotates with respect to the body frame about the $x$ axis they share. This is,
these frames are related by means of a time dependent rotation of $x$ axis and 
angular velocity $\lambda$, in such a way that $\boldsymbol{\omega}$ and 
$\mathbf{L}$ are constant in the precessing reference system, with
values,
\begin{equation}
 \boldsymbol{\omega}_p = \left( \begin{array}{c}
	-\frac{L}{I_1}\sin(\varepsilon) \\ 0 \\
	\frac{L}{I_T}\cos(\varepsilon) \\
     \end{array} \right), 
      \quad \mathbf{L}_p = 
    \left( \begin{array}{c}
     -L\sin(\varepsilon) \\ 0 \\ L\cos(\varepsilon) \\
      \end{array} \right). 
\label{eqn:vel_mom_prec}                                          
\end{equation}

We are interested in the equations of motion of our galactic model in the 
precessing reference frame. To compute them, we need the angular velocity 
of the precessing frame with respect to the inertial one.
This angular velocity, $\boldsymbol{\Omega_p}$, is the sum of the
angular velocity of the body, $\boldsymbol{\omega}_p$, and the angular velocity
of the body axes in the precessing frame. Therefore, and taking into account the 
value of $\lambda$, we obtain,
\begin{equation}
 \boldsymbol{\Omega_p} = \boldsymbol{\omega}_p +\left(\begin{array}{r}
     -\lambda \\ 0 \\ 0 \end{array} \right) =
  \left( \begin{array}{c}
      -\Omega\sin(\varepsilon) \\ 0 \\ \Omega\cos(\varepsilon) \\
    \end{array} \right),
\label{eqn:angvelcte}
\end{equation}
where $\Omega = ||\boldsymbol{\Omega_p}|| = \frac{L}{I_T}$ (see Chapter 2 of 
\citet{patsan} for more details about the rotations). 

Finally, as a complement, we briefly discuss the motion of the bar 
in the inertial frame. According to Poinsot's theorem~\citep[see][]{Arnold}, 
the bar, described by an ellipsoid, rolls without slipping on a fixed 
plane normal to the angular momentum $\mathbf{L}$. If the ellipsoid has axial 
symmetry, as in our case, this motion is the superposition of a rotation of 
the ellipsoid along its symmetry axis with constant angular velocity, 
$\lambda$, and a precession with constant pattern speed $\Omega$ around the 
axis of the angular momentum. The tilt angle, $\varepsilon$, formed by the 
symmetry axis of the ellipsoid and the fixed plane remains constant.

It is also interesting to remark that when in our model we take $\varepsilon=0$,
the misalignment between the angular velocity and the angular momentum 
disappears and we recover the classical model with pattern speed $(0,0,\Omega)$.
\begin{figure}
  \centering
    \includegraphics[width=0.4\textwidth]{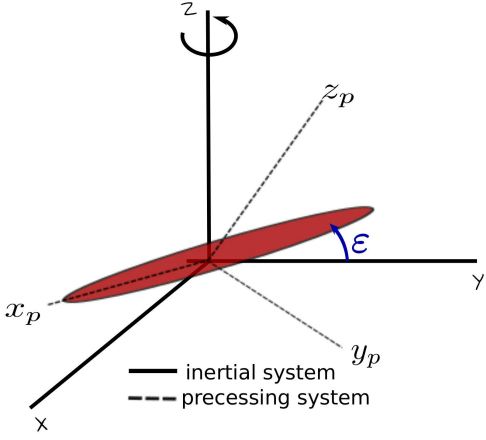}
    \caption{Reference systems. The major axis of the bar is aligned with the precessing $x$ axis and the precessing $z$ axis describes a cone
      about the inertial $Z$ axis. Here $(x_p,y_p,z_p)$ denotes the precessing 
      reference frame and $(X,Y,Z)$ the inertial one.}
      \label{fig:mov_bar}
\end{figure}

We have to emphasize here that this solution is valid for an axially symmetric
rigid body, which in the case of the bar implies that the components \emph{b}
and \emph{c} are equal and therefore the moments of inertia $I_2$ and $I_3$
take the same value. If we want to treat the problem for the triaxial bar
($b\neq c$), the equations of motion~(\ref{eqn:systmodel}) that we present in
the next section would be non-autonomous and this would introduce a 
complexity in the model that is not essential for our research.
We see in the following sections, however, that the behaviour for a bar
with $b\neq c$ is qualitatively the same and that the results for the
axially symmetric and triaxial bar are essentially equivalent.

As for the location of the disc, we have to take into account that we do not
consider that the disc behaves as a rigid body, but it has to follow the main motion
of the bar around the z-axis since the bar has been formed from the disc. So we
include the disc in the equatorial xy plane of the model as a building block.
Thus, essentially the new model provides a gravitational potential once the bar
is formed (like many other traditional barred models
\citep[e.g.]{ContopPapayan,Pfenn,ContopoulosGrosbol}) but generalising the fact that when the
particles rearrange to form the bar, this is not just like a "parallel"
arrangement with the main axes, but it may have small misalignments, that make
the bar precess inside the disc.  Therefore, we can consider the new model
is a perturbation of order $\varepsilon$ of the traditional bar plus
axisymmetric component model. For clear reasons, this fact constrains the value
of $\varepsilon$ forcing it to be small. 

Of course, in a more realistic situation the disc would be somewhat warped, but
we have not included this fact in the model because the small warped amplitude
would not play an essential role in the gravitational potential that is already
considered in the order of the thickness of the disc. On the contrary, a main
point of the precessing model is that, due to symmetries and by construction,
the resulting system is autonomous, making it simpler.

\subsection{Equations of motion associated with the precessing model}

As is well known, the equation of motion of a particle in a rotating system is
\begin{equation}\label{eq:motion_ant}
{\mathbf{\ddot{r}=-\nabla \phi}} -2\mathbf{(\Omega_p \times \dot{r})}-\mathbf{\Omega_p \times (\Omega_p \times r)},
\end{equation}
where ${\bf r}=(x,y,z)$ is the position vector of a star or a particle, $\phi=\phi_{d}+\phi_{b}$ is the potential of the system, which in our case is the sum of the potentials $\phi_{d}$ and $\phi_{b}$ of the disc and bar, respectively, and $\boldsymbol{\Omega_p}$ is the angular velocity given in Eq.~\eqref{eqn:angvelcte}. The second term of the right hand side corresponds to Coriolis acceleration, and the last term to centrifugal acceleration.

To study the trajectories of stars under this
potential, we consider the precessing reference frame as shown in Fig.~\ref{fig:mov_bar}. 

The equations of motion for our precessing reference system, hereafter referred to as the \emph{precessing model}, are obtained by substituting the pattern speed~\eqref{eqn:angvelcte} in Eq.~\eqref{eq:motion_ant}, 
\begin{equation} \label{eq:motion}
  \left\lbrace
  \begin{array}{l}
   \ddot{x}=2\Omega \cos(\varepsilon) \dot{y} + \Omega^2 \cos^2(\varepsilon) x + \Omega^2 \sin(\varepsilon)\cos(\varepsilon)z - \phi_{x} \\
   \ddot{y}=-2\Omega \cos(\varepsilon)\dot{x} -2\Omega \sin(\varepsilon)\dot{z} + \Omega^2 y - \phi_{y} \\
   \ddot{z}=2\Omega \sin(\varepsilon)\dot{y} + \Omega^2 \sin(\varepsilon) \cos(\varepsilon) x + \Omega^2 \sin^2(\varepsilon) z - \phi_{z}\\
  \end{array}
 \right.
\end{equation}
where $\varepsilon$ is the tilt angle, $\Omega$ the modulus of the pattern speed and $\phi$ the potential ($\phi = \phi_{b} + \phi_{d}$).

By setting $(x_1,x_2,x_3,x_4,x_5,x_6)=(x,y,z,\dot{x},\dot{y},\dot{z})$, 
the system~(\ref{eq:motion}) can be written as a system of first order differential
equations, 
\begin{equation}
 \left\lbrace
 \begin{array}{l}
  \dot{x_1} = x_4 \\
  \dot{x_2} = x_5 \\
  \dot{x_3} = x_6 \\
  \dot{x_4} = 2\Omega \cos(\varepsilon) x_5 + \Omega^2 \cos^2(\varepsilon) x_1 + \Omega^2 \sin(\varepsilon)\cos(\varepsilon)x_3 - \phi_{x_1} \\
  \dot{x_5} = -2\Omega \cos(\varepsilon)x_4 -2\Omega \sin(\varepsilon)x_6 + \Omega^2 x_2 - \phi_{x_2} \\
  \dot{x_6} = 2\Omega \sin(\varepsilon)x_5 + \Omega^2 \sin(\varepsilon) \cos(\varepsilon) x_1 + \Omega^2 \sin^2(\varepsilon) x_3 - \phi_{x_3}\\
 \end{array}
 \right.
\label{eqn:systmodel}
\end{equation}
where we recover the classical model when $\varepsilon=0$.
It is also worth mentioning that the precessing 
model has a Jacobi first integral given by,
 \begin{equation}\label{PreCJAC}
 \begin{split}
   & C_J(x_1,x_2,x_3,x_4,x_5,x_6) = -(x_4^2+x_5^2+x_6^2) \\
   &+2\Omega^2\sin(\varepsilon)\cos(\varepsilon)x_1x_3+(\Omega^2\cos^2(\varepsilon)x_1^2\\
   &+\Omega^2x_2^2+\Omega^2\sin^2(\varepsilon)x_3^2)-2\phi,
 \end{split} 
\end{equation}
and the effective potential
is defined by,
\begin{equation}
\begin{split}
 \phi_{_{\hbox{\scriptsize eff}}} &= \phi - \frac{1}{2}\Omega^2 (\cos^2(\varepsilon) x_1^2 + x_2^2 + \sin^2(\varepsilon) x_3^2) \\
 & - \Omega^2\sin(\varepsilon)\cos(\varepsilon)x_1x_3.
 \end{split} 
\end{equation}

The Jacobi integral is related with the Hamiltonian character of the system. 
In fact, and as in the non-precessing case, $C_J=-2H$, where the Hamiltonian in the precessing reference system is,
 \begin{equation}
 \begin{split}
    &H(q_1,q_2,q_3,p_1,p_2,p_3) = \frac{1}{2}(p_1^2+p_2^2+p_3^2) \\
    & +2\Omega\left(\cos(\varepsilon)p_1q_2 -\cos(\varepsilon)p_2q_1 - \sin(\varepsilon)p_2q_3+\sin(\varepsilon)p_3q_2\right) \\
    & +\frac{3}{2}\Omega^2\left(q_2^2+\cos^2(\varepsilon)q_1^2+\sin^2(\varepsilon)q_3^2+2\sin(\varepsilon)\cos(\varepsilon)q_1q_3\right) \\
    & + \phi(q_1,q_2,q_3),
 \end{split} 
\end{equation}
being ($p_i$,$q_i$) the momenta and positions, respectively, in Hamiltonian coordinates defined as
$$
\begin{array}{l}
 q_1 = x_1, \quad  p_1 = \dot{q_1}-2\Omega\cos(\varepsilon)q_2 \, , \\
 q_2 = x_2, \quad  p_2 = \dot{q_2}+2\Omega\cos(\varepsilon)q_1+2\Omega\sin(\varepsilon)q_3 \, ,  \\
 q_3 = x_3, \quad  p_3 = \dot{q_3}-2\Omega\sin(\varepsilon)q_2 \, .
\end{array}
$$

\subsection{The galactic bar potential}

As mentioned above, our galactic model consists of the superposition of an axisymmetric disc plus an ellipsoidal bar.

In this paper and as in \citet{Romero1}, we essentially consider the same 
potential as in~\citet{Pfenn}. The disc component is modelled by a Miyamoto-Nagai potential \citep{Miyamoto},
\begin{equation}
 \phi_d = - \frac{GM_d}{\sqrt{R^2+(A+\sqrt{B^2+z^2})^2}}
\end{equation}
where $R^2=x^2+y^2$ and $z$ denotes the distance in
the out-of-plane component. The parameter G is the gravitational constant and $M_d$ is the mass of 
the disc. The parameters A and B characterise the shape of the disc. 
Parameter A measures the radial scale length of the disc while B is a measure of 
the disc thickness in the $z$ direction. Since galactic discs are larger in 
the radial direction than in the vertical one, A is greater 
than B.

The barlike part is modelled by a Ferrers ellipsoid \citep{Ferrers},
with a density function,
\begin{equation}
 \rho = 
 \left\lbrace
 \begin{array}{l}
  \rho_0(1-m^2)^{n_h}, \hspace{1cm} m\leq 1 \\
  0,  \hspace{2.8cm} m > 1 \\
 \end{array}
 \right.
\end{equation}
where $m^2=x^2/a^2 + y^2/b^2 + z^2/c^2$. 
The parameters \emph{a} (semi-major axis) and \emph{b}, \emph{c}  (intermediate and semi-minor axes) 
determine the shape of the bar, parameter \emph{n$_h$} determines the homogeneity degree for the
mass distribution, $\rho_0 = \frac{105}{32 \pi}\frac{GM_b}{abc}$ is the central density if $n_h=2$, and $M_b$ the mass of the bar. This model concentrates matter in the central region and decreases smoothly towards zero at a finite distance. 

The density of the bar potential is related with its potential, $\phi_{b}$, by means of 
the Poisson equation ($\nabla^2 \phi = 4 \pi G \rho$):
\begin{equation}
 \phi_b=-\pi \,G\, a b c \frac{\rho_0}{n_h+1}\int^{\infty}_{\bar{\lambda}}\frac{du}{\sqrt{\Delta(u)}}(1-m^2(u))^{(n_h+1)},
\end{equation}
where G is the gravitational constant, $\Delta(u)=(a^2+u)(b^2+u)(c^2+u)$ 
and $\bar{\lambda}$ the unique positive solution of $m^2(\bar{\lambda})=1$ 
if $m\geq 1$, 
(that is, if the particle lies outside the bar), and zero otherwise.

The length unit used throughout this work is the kpc, the time unit is u$_t = 2 \times 10^6$ yr and the gravitational
constant $G = 6.674 \times 10^{-11}$ m$^3$ kg$^{-1}$ s$^{-2}$. 
In this paper we take values $A=3$, $B=1$ for the disc, 
and for the bar we are going to consider two different Ferrers bars, one symmetric and another one with the values taken by \citet{Pfenn}. For both bars, the homogeneity index is set to $n_h=2$ and the semi-major axis of the bar to $a=6$. Whereas the first bar has revolution symmetry with semi-minor axes $b=c=0.95$, the second bar has just axial symmetry with $b=1.5 \neq c=0.6$.
Some other parameters are considered to vary within a range: 
$GM_d \in [0.6, 0.9]$, $GM_b \in [0.1,0.4]$ 
(but having in mind that 
$G(M_d+M_b)=1$). Finally we also consider the pattern speed 
$\Omega \in [0.05,0.06]$~[u$_t$]$^{-1}$ ($\approx [24.46,29.36]$~km/s/kpc) 
and the tilt angle 
$\varepsilon \in [0,0.2]$ rad = $[0, 11{.}46]^{\circ}$. 

A useful property in order to study the matter distribution in galactic models is the rotation curve or circular-speed $V_{rot}(r)$, defined as the speed of a particle of negligible mass in a circular orbit at radius $r$. For a potential $\phi$, we define $V_{rot}$ as
\begin{equation}
 V_{rot}^2 = r \frac{d\phi}{dr} \, .
\end{equation}   

\begin{figure}
    \centering
     \includegraphics[width=0.4\textwidth]{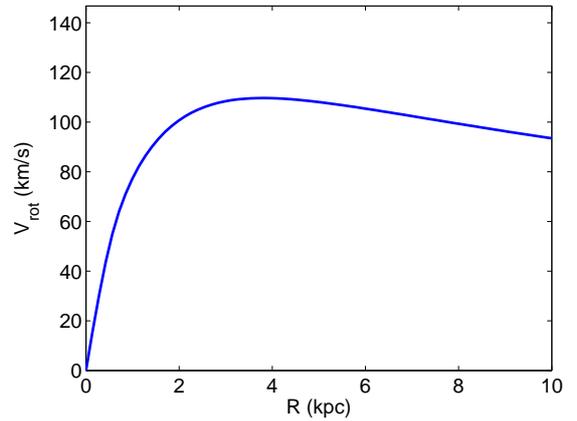}
     \caption{Rotation curve of the potential $\phi = \phi_b+\phi_d$.}
   \label{fig:rot_curv}
\end{figure}

Although we have not imposed a halo potential, the rotation curve of the total potential, composed by the Ferrers bar and the Miyamoto-Nagai disc, is rather flat in the outer parts (see Fig.~\ref{fig:rot_curv}).

\subsection{Characteristics of the precessing model}
\begin{figure}
  \centering
    \includegraphics[width=0.45\textwidth]{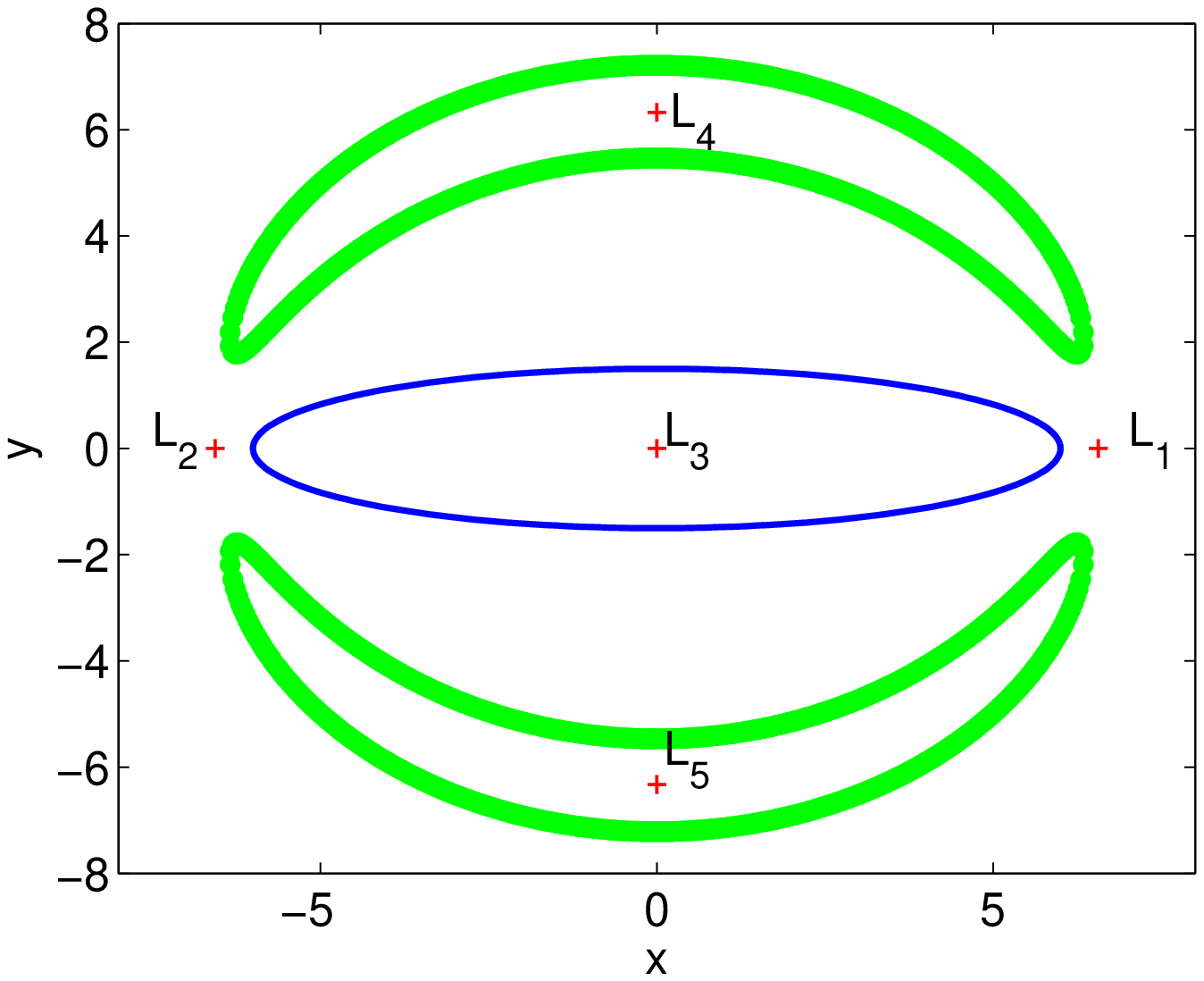}    
    \includegraphics[width=0.45\textwidth]{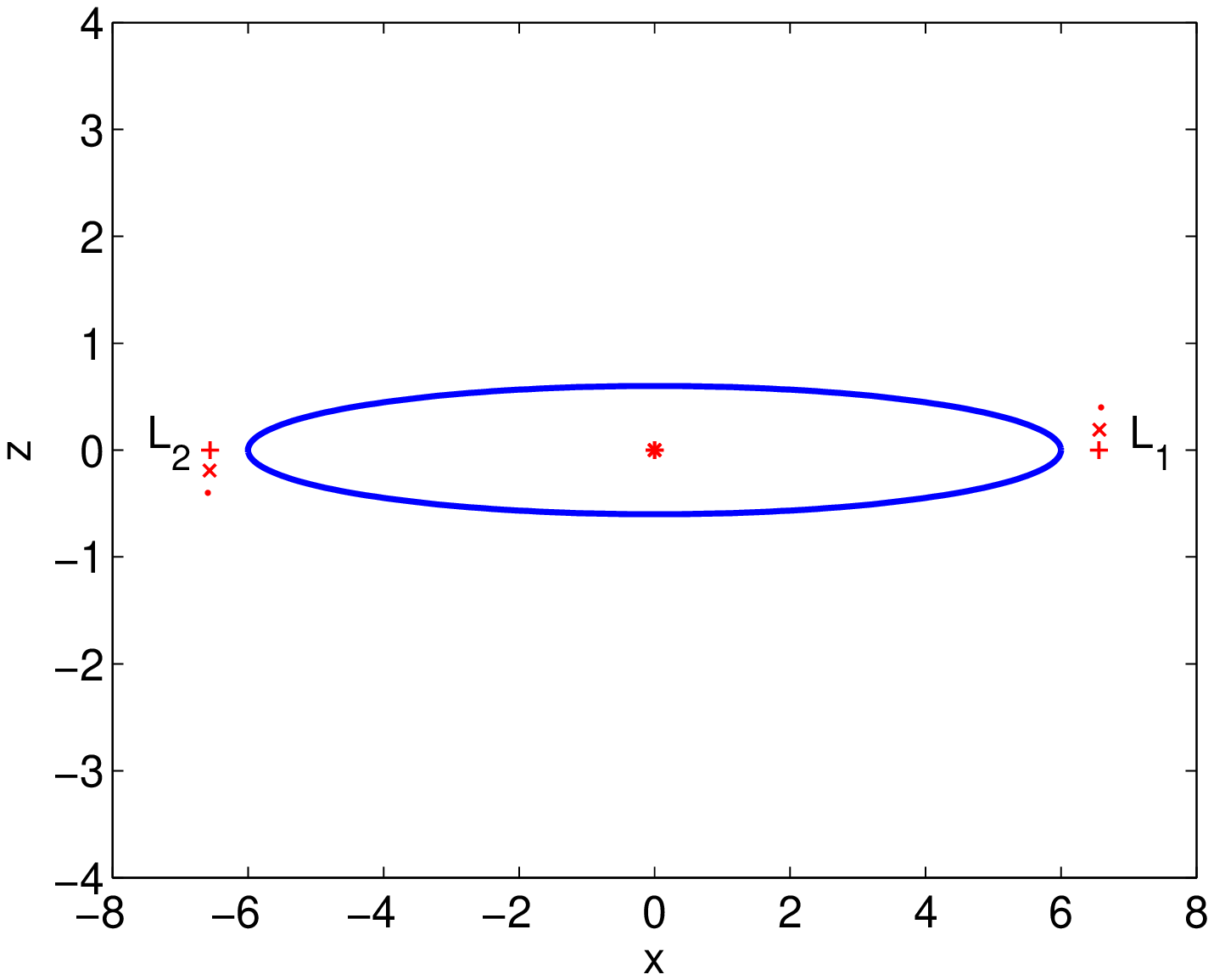}
  \caption{Equilibrium points of the precessing model with $GM_b=0.1$. Top: $xy$ plane with zero velocity curves for a Jacobi constant $C_J=-0.1876$. Bottom: $xz$ plane,'+' for $\varepsilon=0$, '$\times$' for $\varepsilon=0{.}1$, '{\tiny$\bullet$}' for $\varepsilon=0{.}2$, while the '$\ast$' shows the position of the central $L_3$ point. The blue curve in both panels outlines the triaxial Ferrers bar.}
  \label{fig:peq}
\end{figure}

We define zero velocity surfaces of the precessing model as the manifold
 $(x_1,x_2,x_3)\in \mathbb{R}^3$ defined by Equation~(\ref{PreCJAC}) with $x_4=x_5=x_6=0$
for a given value of the Jacobi integral $C_J$.
Their cut with the $z=0$ plane define zero velocity curves and the regions 
where $\phi_{_{\hbox{\scriptsize eff}}}>C_J$ are forbidden regions for a star 
of the given energy (see Fig.~\ref{fig:peq}).

\begin{figure}
 \centering
  \includegraphics[width=0.5\textwidth]{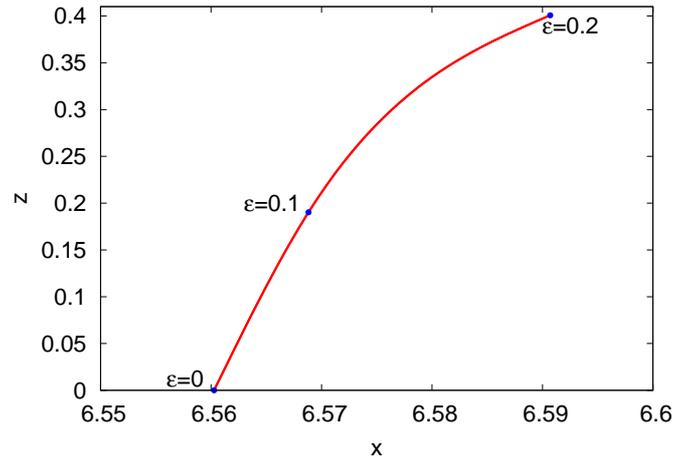}
  \caption{Variation of the $xz$ coordinates of $L_1$ as $\varepsilon$ varies within the range $\varepsilon \in [0,0{.}2]$ for the precessing model with $GM_b=0.1$. The marked dots show the $L_1$ coordinates for three specific values of $\varepsilon$: $\varepsilon=0$ (bottom part of the plot) $\varepsilon=0.1$ (middle) and $\varepsilon=0.2$ (top).}
  \label{fig:L1_eps}
\end{figure}

As in the case $\varepsilon =0$, our precessing model in rotating coordinates 
has five Lagrangian equilibrium points ($L_i$, $i=1\ldots 5$), solutions of 
${\bf \nabla} \phi_{\hbox{\scriptsize eff}} = 0$. 
These are represented in Figs.~\ref{fig:peq} and \ref{fig:L1_eps}. As for the 
properties of these libration points when $\varepsilon=0$, $L_1$ and $L_2$ lie 
on the $x$-axis and are symmetric with respect to the origin. 
$L_3$ lies on the origin of coordinates, and, $L_4$ and $L_5$ lie on the 
$y$-axis and they are also symmetric with respect to the origin (see 
the upper panel of Fig.~\ref{fig:peq}). 
The two equilibrium points, $L_1$ and $L_2$ are unstable while $L_3$ is 
linearly stable and it is surrounded by the $x_1$ family of periodic orbits 
which is responsible of maintaining the bar structure, while the stable points 
$L_4$ and $L_5$ when $\varepsilon=0$ have been thoroughly examined and are surrounded by families of periodic banana orbits 
\citep{Athan1983, Contop1981, Skokos}. 

In our precessing model we notice that whereas $L_3$, $L_4$, $L_5$ maintain 
their coordinates fixed independently of $\varepsilon$, $L_1$ and $L_2$ vary 
as $\varepsilon$ changes. Of relevant importance is the out-of-plane 
$z$-component for $L_1$ and $L_2$ (see bottom panel of Fig.~\ref{fig:peq}). 
In Fig.~\ref{fig:L1_eps} we detail the evolution of the coordinates of the 
equilibrium point $L_1$ when the parameter $\varepsilon$ varies from 
$\varepsilon=0$ to $\varepsilon=0.2$.

\section{The structure of periodic orbits inside the bar} \label{sect:section3}

In order to understand the formation, evolution and properties of any given 
structure it is essential to first understand its building blocks. In the case 
of galactic dynamics, and particularly for barred galaxies, it has been 
clearly demonstrated that some building blocks are periodic orbits elongated 
along the bar. The study of these building blocks provided answers to a number 
of crucial questions, like why bars are bisymmetric, why they rotate as rigid
bodies, why they can not extend beyond corotation, etc. 
\citep[see][among others]{Contop1981, Athan1983, Pfenn, Skokos}.

In this section we start from the infinitesimal periodic orbits about the
central equilibrium point $L_3$ and considering either the period or the
energy, we continue the families of periodic orbits inside the bar at the same
time we study their stability properties. We obtain therefore evidence that the
stable orbits we find give structure to the bar, since stars or particles can
be trapped in their neighbourhood.  
All the integrations in this work have been carried out numerically using
a Runge-Kutta-Fehlberg method of orders 7-8. This method not
only assures the conservation of the Jacobi constant, but it also provides the
required accuracy for the detection of periodic orbits in the dynamical
system.

Proceeding with this aim, we first analyse the stability of the equilibrium 
point $L_3$ in the precessing model. Let us consider the differential matrix around any Lagrangian 
point of the system (\ref{eqn:systmodel}): 
\begin{equation}
DF(L_i) =
 \left( \begin{array}{cccccc}
  0 & 0 & 0 & 1 & 0 & 0 \\
  0 & 0 & 0 & 0 & 1 & 0 \\
  0 & 0 & 0 & 0 & 0 & 1 \\
  a & \phi_{x_1x_2} & b & 0 & c & 0 \\
  \phi_{x_1x_2} & d & \phi_{x_2x_3} & -c & 0 & -e \\
  b & \phi_{x_2x_3} & f & 0 & e & 0
 \end{array} \right)_{L_i}
\label{eqn:df}
\end{equation}
where 
\begin{equation}
 \left\lbrace
 \begin{array}{l}
 a = \Omega^2\cos(\varepsilon)^2+\phi_{x_1x_2}, \\
 b = \Omega^2\sin(\varepsilon)\cos(\varepsilon)+\phi_{x_1x_3}, \\
 c = 2\Omega\cos(\varepsilon), \\
 d = \Omega^2+\phi_{x_2x_2}, \\
 e = 2\Omega\sin(\varepsilon), \\
 f = \Omega^2\sin(\varepsilon)^2+\phi_{x_3x_3}.
 \end{array}
 \right.
\end{equation}

For the particular case of $L_3$, the eigenvalues of (\ref{eqn:df}) are of 
the form $\{\lambda i$, $-\lambda i$, $\mu i$, $-\mu i$, $\omega i$, 
$-\omega i\}$ ($\lambda$, $\mu$, $\omega$ $\in \mathbb{R}^+$), for any 
selected value of $\varepsilon$. Since the purely imaginary eigenvalues are
associated to infinitesimal librations, the linearised flow around $L_3$ in 
the rotating frame of coordinates is characterised by a superposition of 
three oscillations. This is, the $L_3$ Lagrangian point is a linearly stable elliptic point and in Dynamical Systems this behaviour is usually denoted in
the form centre$\times$centre$\times$centre. In our case it has two centre components
inside the $xy$ plane and another one in the $z$ direction.

Next, following 
the work of \citet{Pfenn}, which is a particular case of the precessing model with 
$\varepsilon=0$, and following \citet{Broucke}, \citet{Hadjidemetriou}, we define the stability indexes of the periodic orbits, $b_1$, $b_2$:
\begin{equation}
  b_1=-(\lambda+1/\lambda),\quad b_2=-(\mu+1/\mu).
\end{equation}
With these definitions, a periodic orbit is stable only when $b_1$ and $b_2$ are real 
and $|b_1|,|b_2| \lid 2$, otherwise it is unstable. If $|b_1|$ or $|b_2|=2$, the Jacobian matrix of the continuation process is degenerate and 
bifurcations of the family are allowed. If $b_i=+2$, the bifurcation occurs 
through period doubling, while if $b_i=-2$, the bifurcation keeps the same 
period.

\begin{figure}
    \vspace{-0.2cm}
    \hspace{-0.7cm}\includegraphics[width=0.6\textwidth, angle=0]{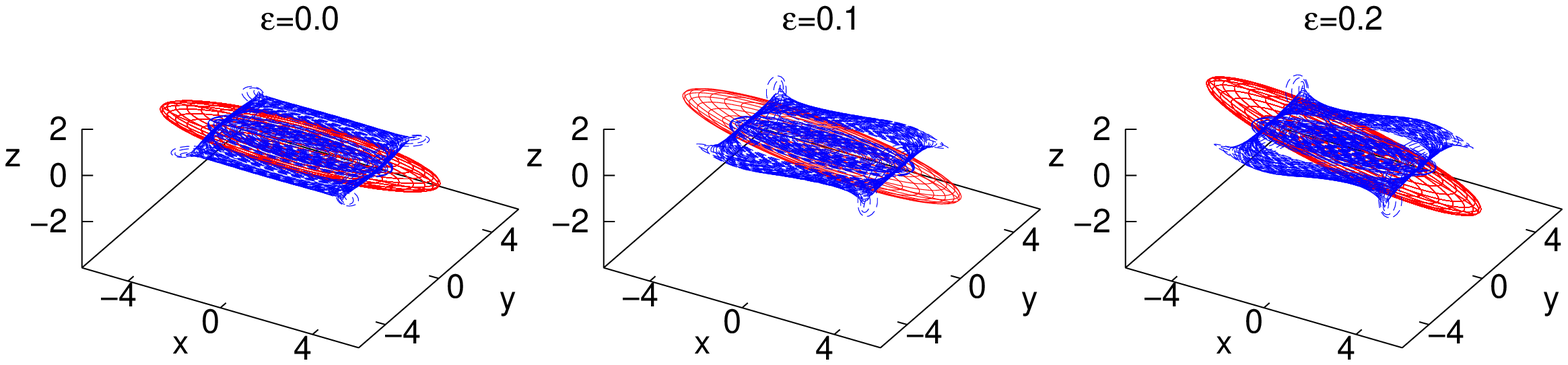} 
    \hspace{-1cm}\includegraphics[width=0.5\textwidth, angle=0]{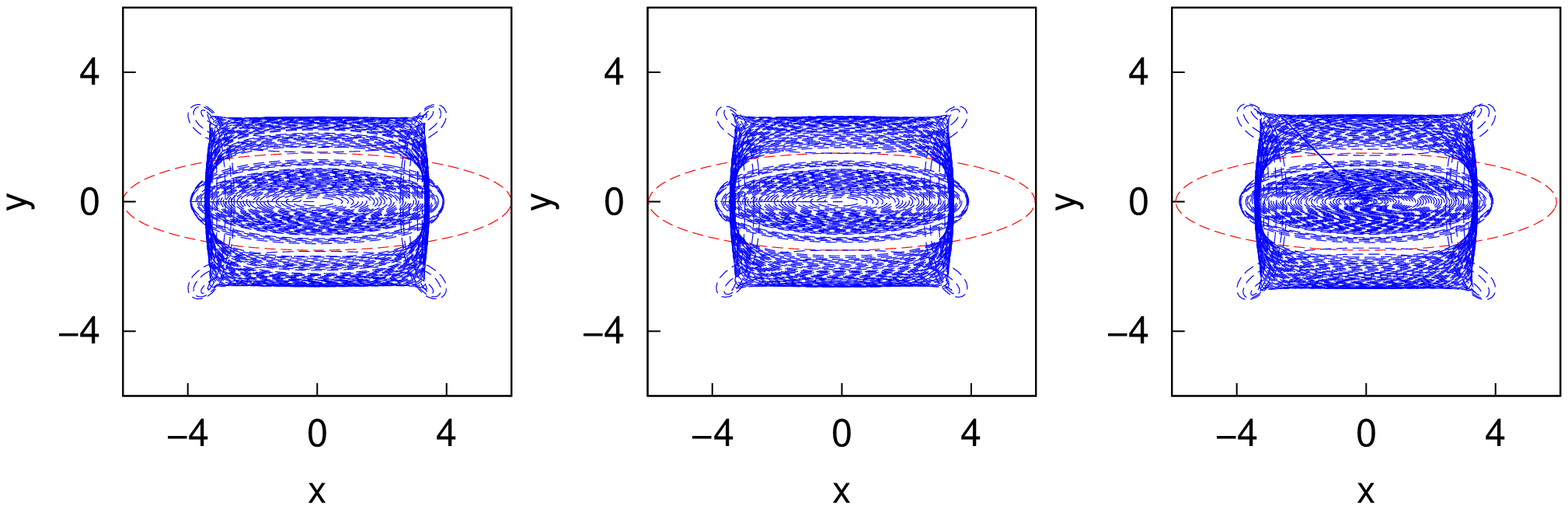} 
    \hspace{-1cm}\includegraphics[width=0.5\textwidth, angle=0]{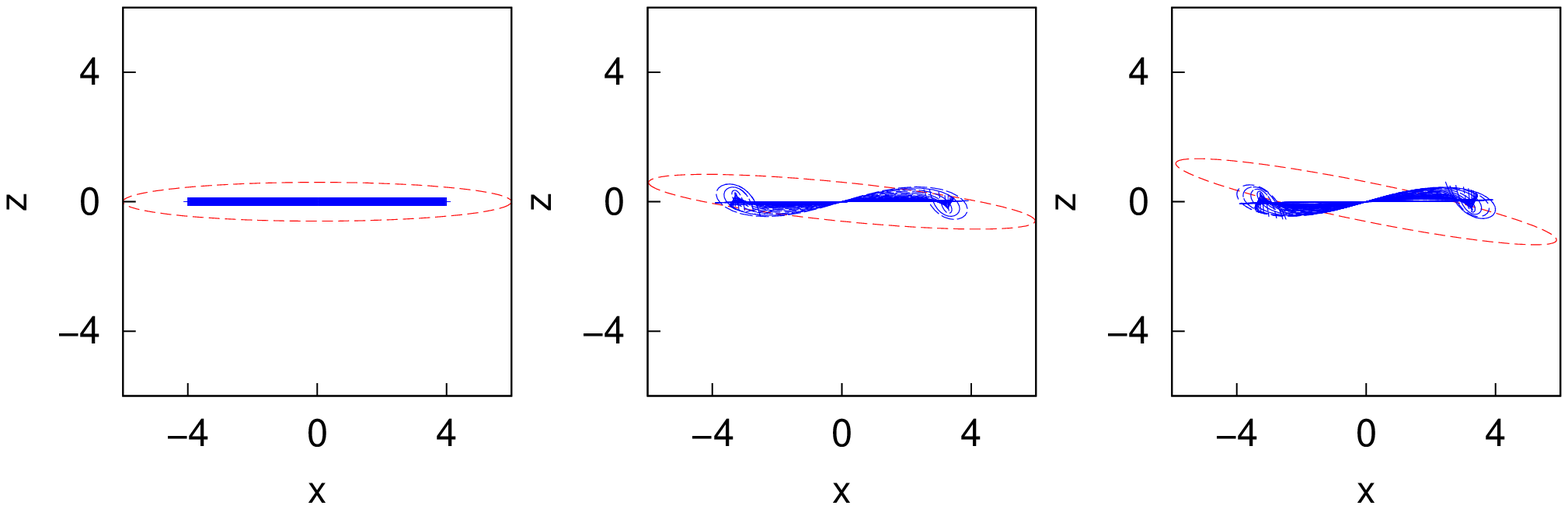} 
    \hspace{-1cm}\includegraphics[width=0.5\textwidth, angle=0]{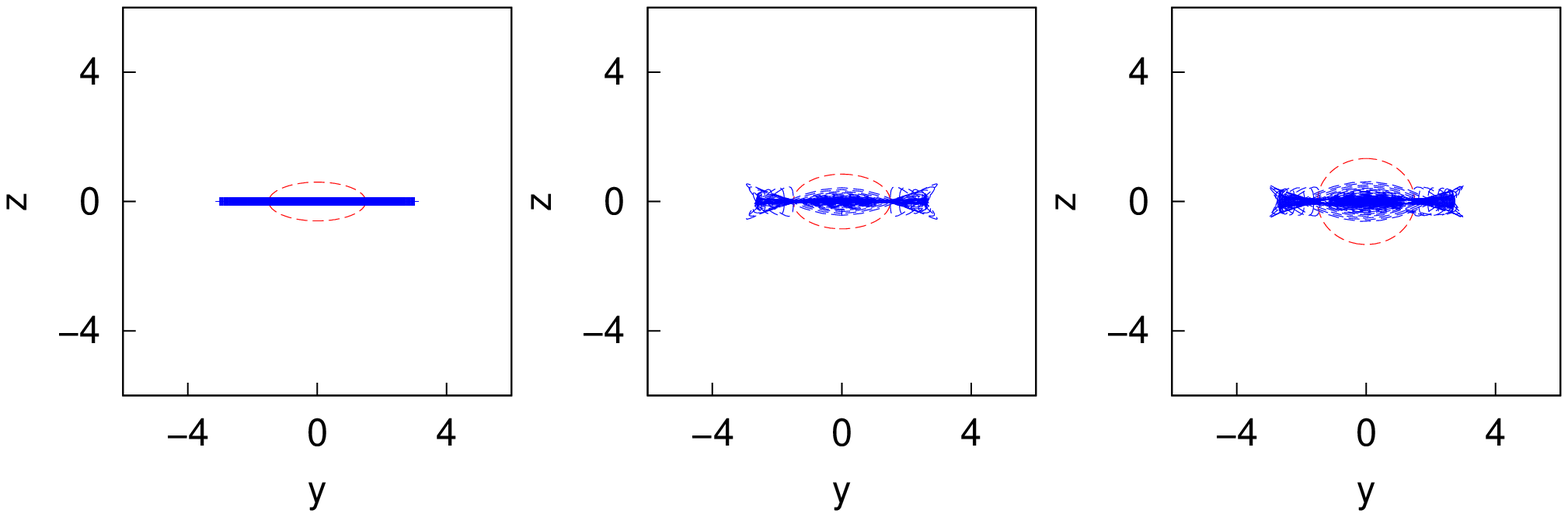} 
    \hspace{-1cm}\includegraphics[width=0.5\textwidth, angle=0]{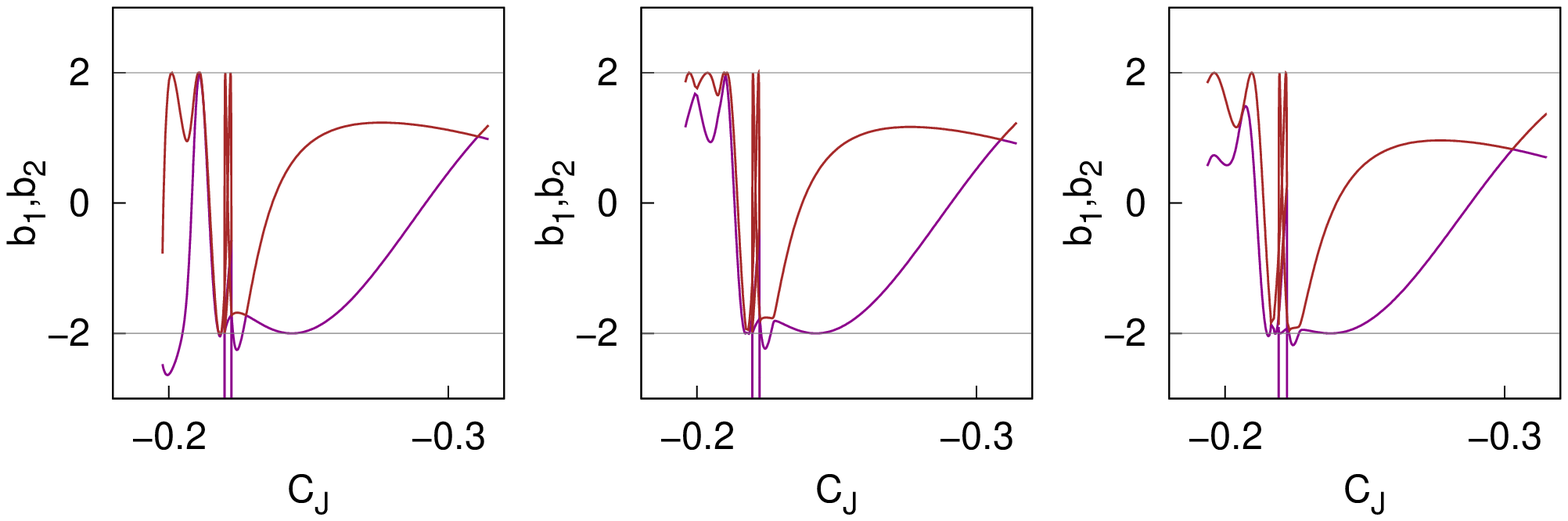} 
  \caption{Family of periodic orbits of the precessing model with $b\ne c$ and $GM_b=0.1$ (first row: 3D view; second, third and fourth rows: (x,y), (x,z) and (y,z) projections, respectively) and stability indexes (bottom row) for $\varepsilon=0,0.1,0.2$ (left, middle and right columns, respectively). In the first four rows, the red dashed lines outline the position of the triaxial Ferrers bar, while the blue lines are periodic orbits of the precessing model. In the bottom row: dark magenta (brown) lines mark the $b_1$ ($b_2$) stability index.}
  \label{fig:mult_int}
\end{figure}

In the left panel of Fig.~\ref{fig:mult_int} we show the results obtained for $GM_b=0.1, \Omega=0.05$ and $\varepsilon=0$, which, for reference, appears in Fig. 4 of \citet{Pfenn}. In this figure, red dashed lines show the contour of the bar with semi-axis $a=6, \, b=1.5, \, c=0.6$  in each plane, whereas blue lines indicate the x1 family of periodic orbits of the model and its bifurcations. 

In a two dimensional model the x1 family of planar periodic orbits about L$_3$ is mainly stable and has been regarded as responsible for the skeleton of the bar's structure. But, in three dimensional models, the backbone of the bars is the x1 family together with its 3D bifurcating families \citep{Skokos, Skokosb}. A 
continuation process in $\varepsilon$ is then used to obtain parallel 
results for the tilt angle $\varepsilon \neq 0$.

\begin{figure}
  \centering
    \includegraphics[width=0.3\textwidth]{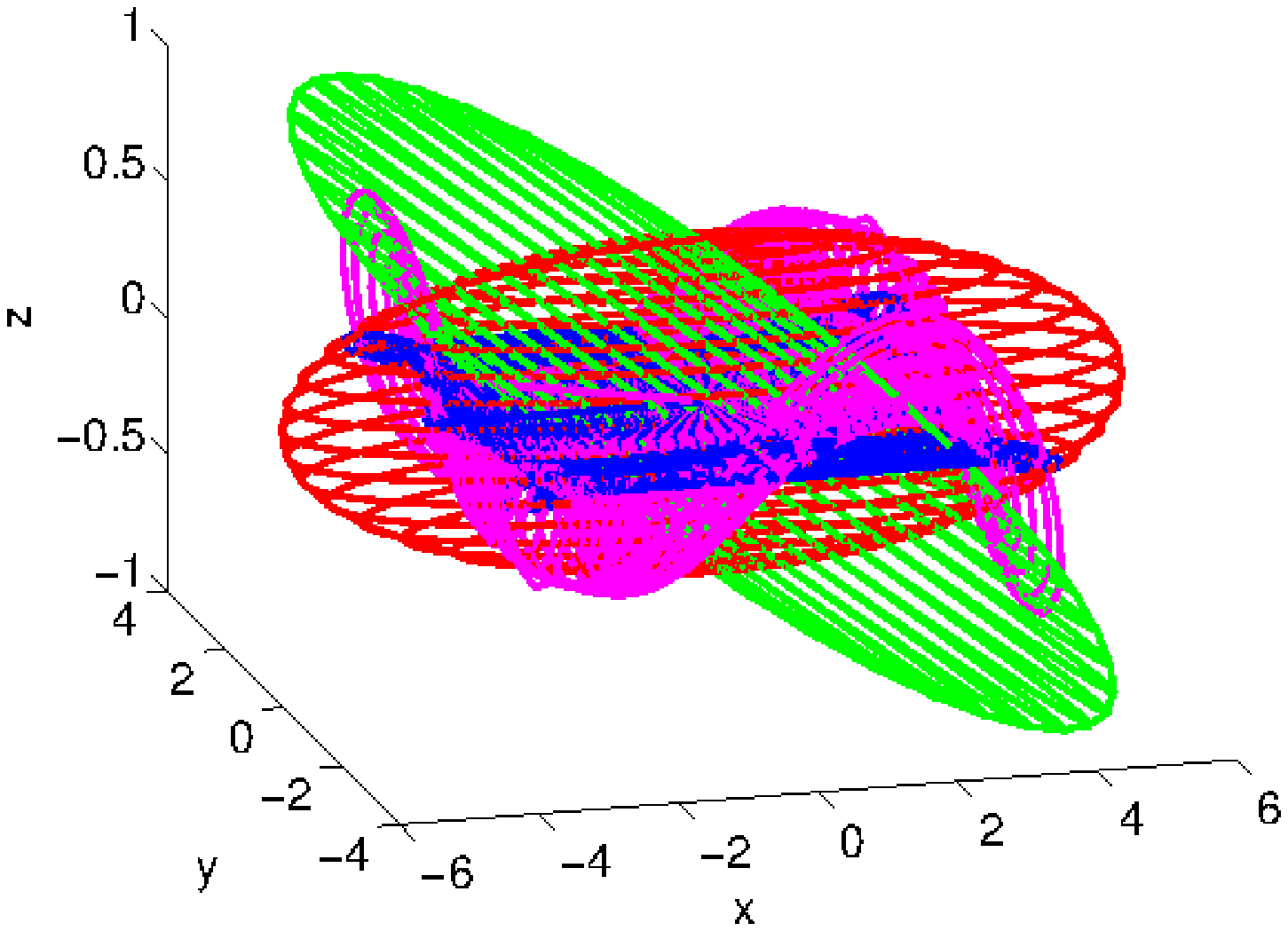}
    \includegraphics[width=0.2\textwidth]{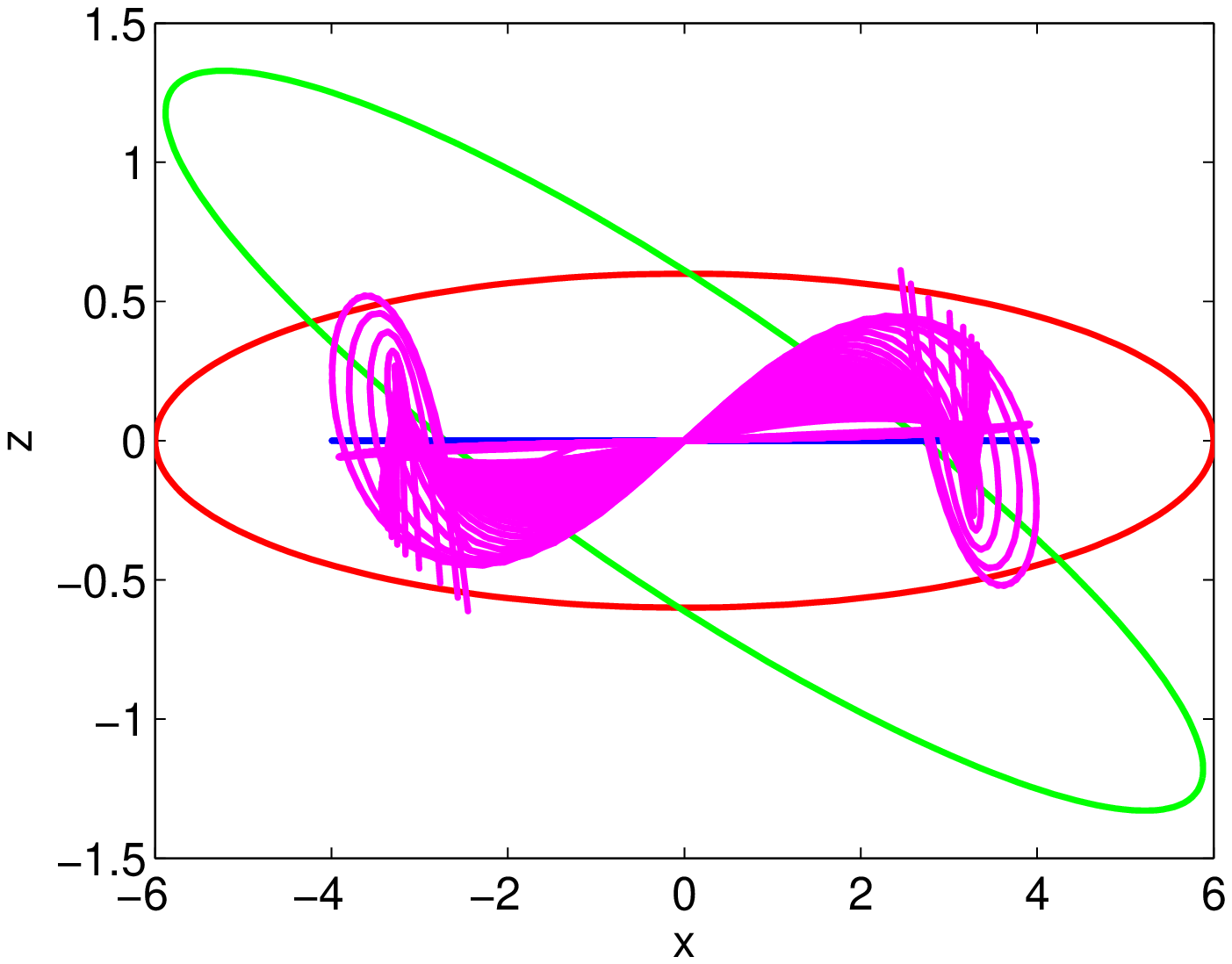}
    \includegraphics[width=0.2\textwidth]{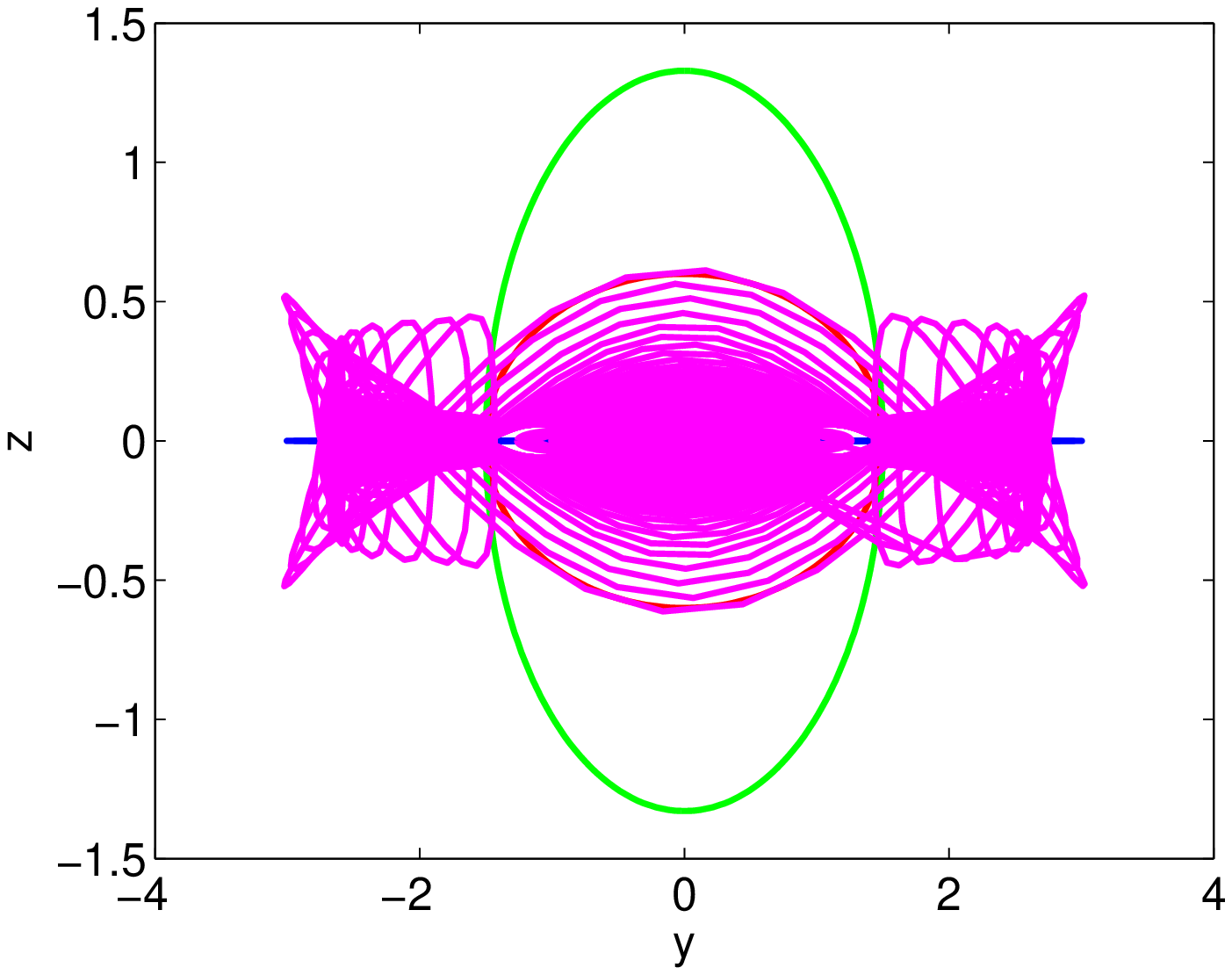}
  \caption{Family of periodic orbits of the precessing model with $b\ne c$ for $\varepsilon=0$ (blue) and $\varepsilon=0{.}2$ (purple) and $GM_b=0.1$. From top to bottom: 3D view, xz-plane (left), yz-plane (right). Red (green) lines show the position of the bar for $\varepsilon=0$ ($\varepsilon=0.2$). Blue (purple) lines indicate the periodic orbits of the model for $\varepsilon=0$ ($\varepsilon=0.2$).}
  \label{fig:int_eps0y02}
\end{figure}

The results obtained in continuing the x1 family and its bifurcations for $\varepsilon =0.1$ and $\varepsilon =0.2$ are shown in the middle and right panels of
Fig.~\ref{fig:mult_int}. Note that, due to the nature of the tilting, the most significant 
change is in the $z$ component. The $xy$ projections remain essentially
the same and the families of periodic orbits continue giving structure to the
bar, as is displayed in Fig.~\ref{fig:int_eps0y02} where the families for
$\varepsilon =0$ and $\varepsilon =0.2$ are compared. In this figure, red (green) lines show the position of the bar for $\varepsilon=0$ ($\varepsilon=0.2$). Blue (purple) lines indicate the periodic orbits of the model for $\varepsilon=0$ ($\varepsilon=0.2$).

In the last row of Fig.~\ref{fig:mult_int}
we compare the stability indexes. For a given value of $\varepsilon$, $b_1$ and $b_2$ cross the 
limits ($\pm 2$) an equal number of times and approximately at the same 
value of the Jacobi constant, C$_J$. All these facts reinforce again the evidence that the 
families of periodic orbits about $L_3$ for any $\varepsilon$ are 
qualitatively the same.

\begin{figure}
    \vspace{-0.2cm}
    \hspace{-0.7cm}\includegraphics[width=0.6\textwidth, angle=0]{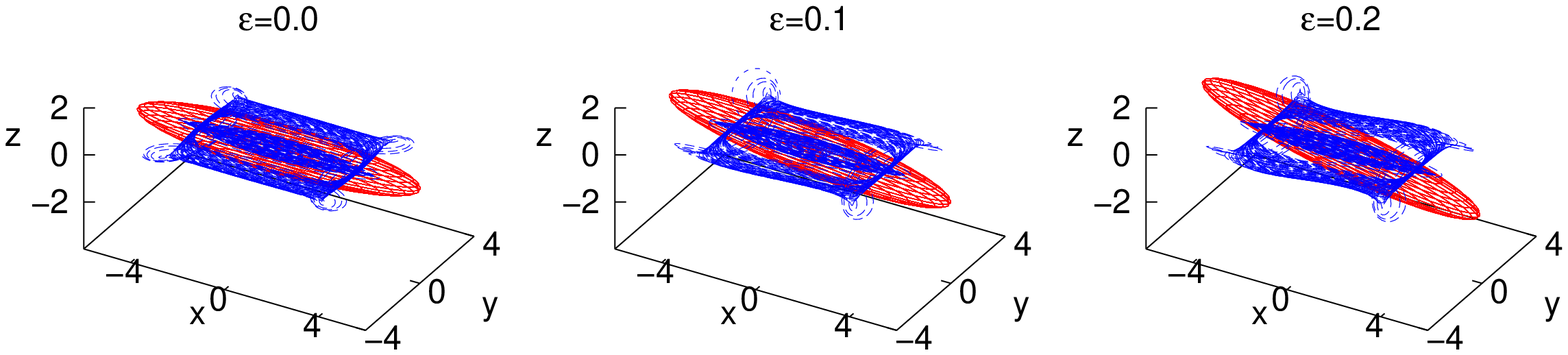} 
    \hspace{-1cm}\includegraphics[width=0.5\textwidth, angle=0]{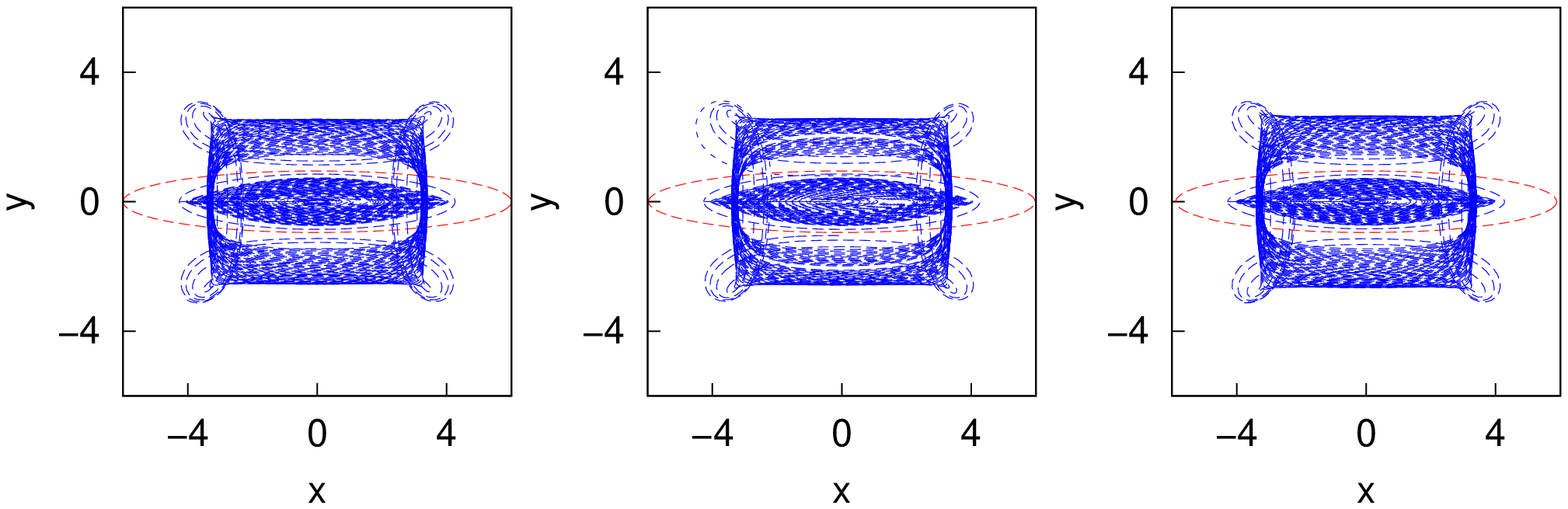} 
    \hspace{-1cm}\includegraphics[width=0.5\textwidth, angle=0]{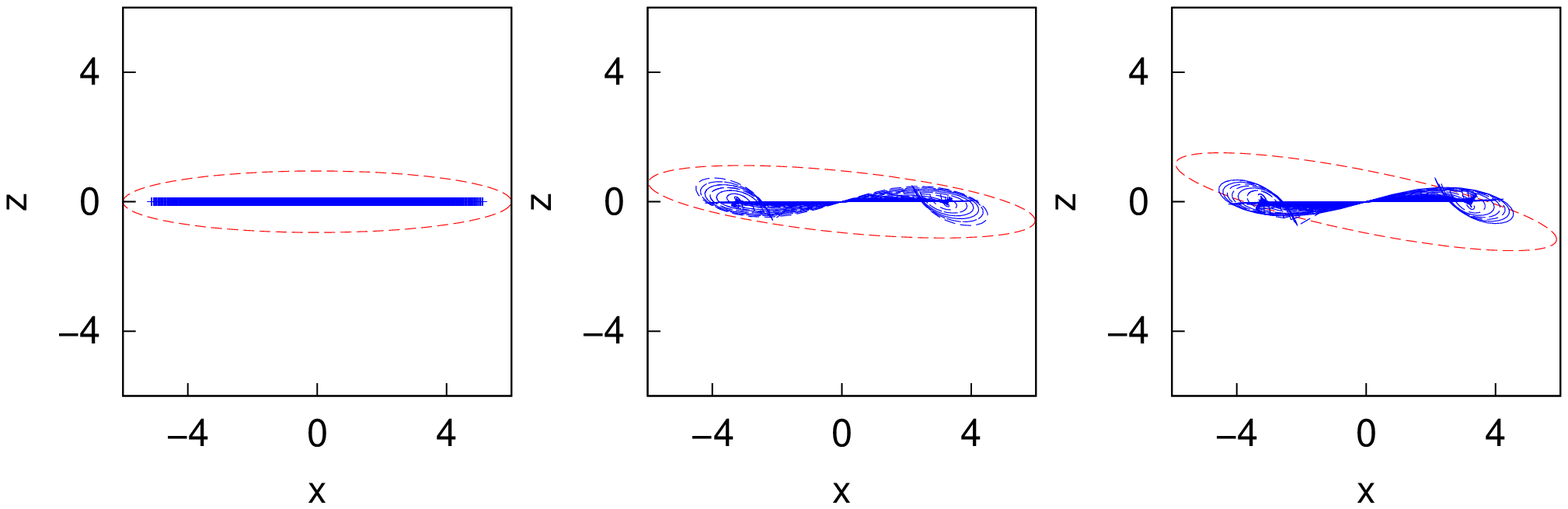} 
    \hspace{-1cm}\includegraphics[width=0.5\textwidth, angle=0]{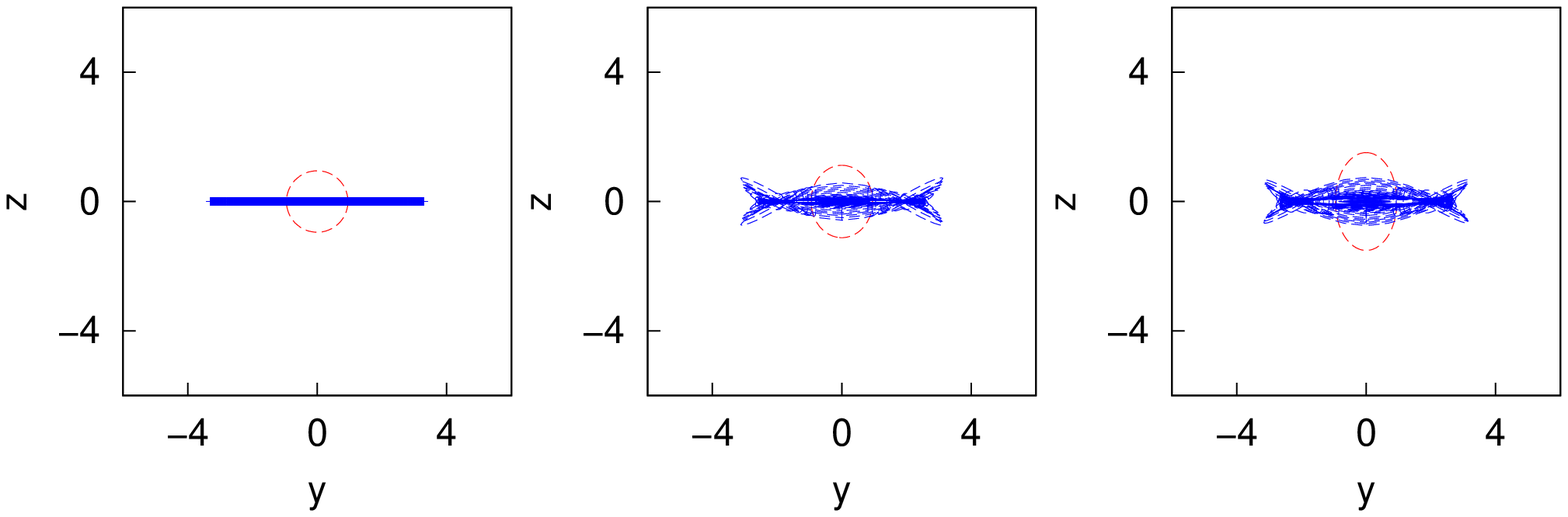} 
    \hspace{-1cm}\includegraphics[width=0.5\textwidth, angle=0]{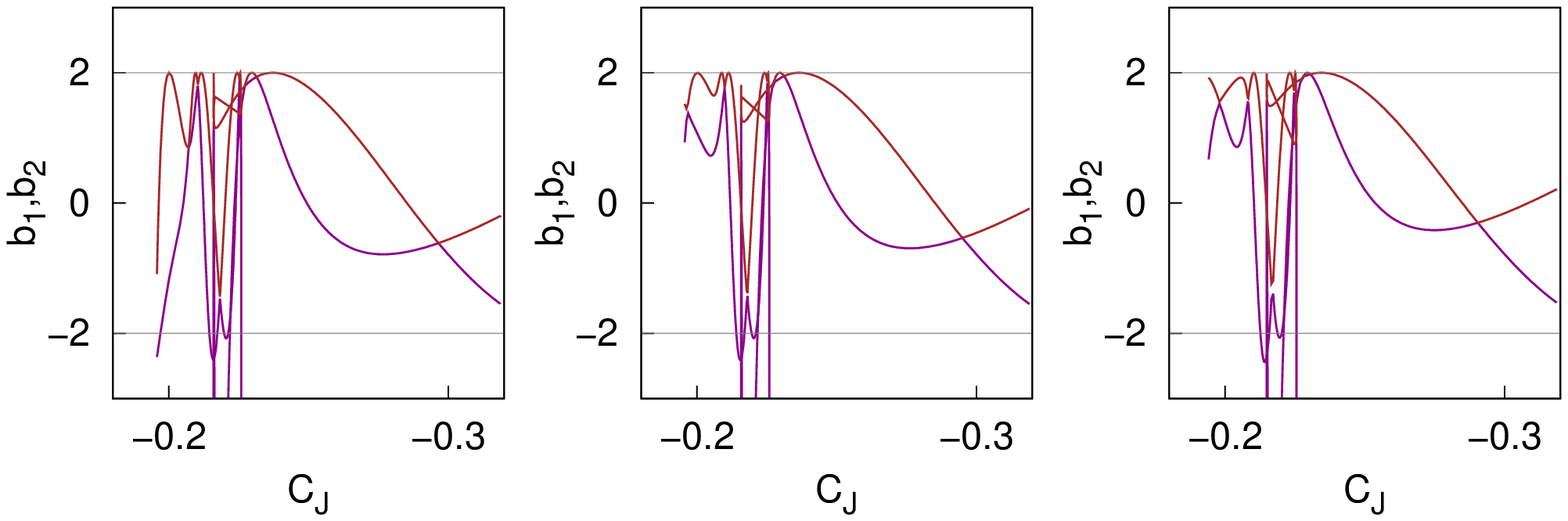} 
  \caption{Family of periodic orbits of the precessing model with $b=c=0.95$ and $GM_b=0.1$ (first row: 3D view; second, third and fourth rows: (x,y), (x,z) and (y,z) projections, respectively) and stability indexes (bottom row) for $\varepsilon=0,0.1,0.2$ (left, middle and right columns, respectively). In the first four rows, the red dashed lines outline the position of the triaxial Ferrers bar, while the blue lines are periodic orbits of the precessing model. In the bottom row: dark magenta (brown) lines mark the $b_1$ ($b_2$) stability index.}
  \label{fig:mult_int_bc}
\end{figure}

At this moment, we can prove that although the equations given in Section~\ref{sect:section2} are for the case in which $b=c$ in the bar, the results given in this section for the axially symmetric model remain the same. In order to evidence this statement, we show in Fig.~\ref{fig:mult_int_bc} how the periodic orbits and the stability indexes remain unchanged for $b=c$ (following the same colour convention as in Fig.~\ref{fig:mult_int}). Note that in order to compare the symmetric case ($b=c$) with the non-symmetric one (model given in~\citet{Pfenn} with $b=1.5 \neq c=0.6$), we impose in both models equal bar mass ($GM_b=0.1$), equal homogeneity ($n_h=2$) and therefore equal particle distribution. Thus, we take a symmetric bar where the parameters $\tilde{b}$ and $\tilde{c}$ are the geometric
mean of the previous parameters, i.e. $\tilde{b}= \tilde{c}=\sqrt{b\cdot c} = 0.95$. This results, somehow, in a gravitational field which
is the average along time of the previous one.

Comparing Figs.~\ref{fig:mult_int} and ~\ref{fig:mult_int_bc}, we can confirm that the family of periodic orbits around the central equilibrium point $L_3$ remain essentially the same, independently of whether we use a symmetric bar or not. Moreover, observing the stability indexes for both models (bottom panels), periodic orbits are in a comparable range of values of the Jacobi constant, being the cuts of the indexes within the limits $|\pm 2|$ qualitatively equal. 

Therefore, we can conclude this comparison saying that in both models,
namely the one with revolution symmetry and the one axially symmetric, the periodic orbits are responsible for maintaining the structure of the bar and giving consistency to the model. In this way, we could use any of both models, but since a bar with parameters $b\neq c$ is more commonly used and we want to compare with the model given in~\citet{Pfenn} to prove that our model is consistent although we apply a tilt, we prefer to show the rest of results for $b \neq c$.

\section{The invariant manifolds in the precessing model} \label{sect:section4}

Once we have analysed the behaviour of periodic orbits inside the bar, we 
continue the study considering trajectories outside the bar that are 
responsible for the main visible building blocks in the barred galaxies, 
like spirals and rings, i.e. the normally hyperbolic invariant manifolds 
associated to the libration point orbits 
about $L_1$ and $L_2$. The set of these
orbits is responsible for the transport of matter between the neighbourhood
of the bar and the exterior part of the galaxy. The stars trapped in these
manifolds make visible their structure in the form of rings and arms. Note that the rings and spirals obtained are response rings and spirals from the bar potential, they are not imposed in the galactic potential, that is, they are not self-gravitating.

We study now the stability character of the libration points L$_1$ and L$_2$.
The eigenvalues of the differential matrix (\ref{eqn:df}) around $L_1$ and 
$L_2$ are of the form $\{\lambda$, $-\lambda$, $\mu i$, $-\mu i$, $\omega i$, 
$-\omega i\}$ ($\lambda$, $\mu$, $\omega$ $\in \mathbb{R}^+$), for any value 
of $\varepsilon$. This is, the two real eigenvalues are related to a 
hyperbolic behaviour like a saddle, whereas the purely imaginary are 
associated to libration motions. This implies that the linearised flow around 
$L_1$ and $L_2$ in the rotating frame of coordinates is characterised by a 
superposition of a saddle and two harmonic oscillations and in Dynamical
Systems this is usually described as a saddle$\times$centre$\times$centre
behaviour. Then $L_1$ and $L_2$ are unstable and are called hyperbolic points.
The dynamics around the unstable equilibrium points in our context are
described in detail in \citet{Romero1} and \citet{CanMas}. Here just a brief
summary follows.

\begin{figure}
   \hspace{-0.3cm} \includegraphics[width=0.5\textwidth,angle=0]{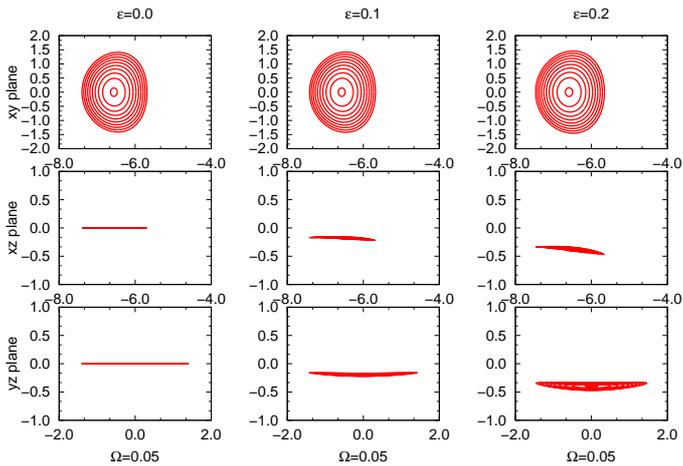}
    \caption{Lyapunov family of periodic orbits around L$_2$ for the model with $GM_b=0.1$ and for a range of values of the Jacobi constant in $(C_{J,L_2},C_{J,L_2}+2\times 10^{-3})$, where $C_{J,L_2}(\varepsilon =0)=-0.1879$, $C_{J,L_2}(\varepsilon =0.1)=-0.1876$ and $C_{J,L_2}(\varepsilon =0.2)=-0.1865$. The tilt angle $\varepsilon \in [0,0.2]$. Note the varying scale of the vertical axis.}
  \label{fig:lyap}
\end{figure}

As is well known, around each unstable equilibrium point, $L_1$ and $L_2$, 
there must exist a family of periodic orbits associated with the eigenvalues
of the elliptical part. They are the planar and vertical families of Lyapunov 
periodic orbits. These orbits are unstable in the vicinity of the equilibrium
point. The vertical family of Lyapunov orbits has been computed in galactic potentials \citep[e.g.][]{Oll98,Rom09} and its structure is different from the planar family shown in Fig.~\ref{fig:lyap}. The vertical family extends to both sides of the galactic plane, while the planar family in the precessing model when the parameter $\varepsilon \neq 0$ remains in one side of the galactic plane, without crossing it.
Furthermore, in \citet{Rom09} it is shown that the family relevant to the transfer of matter within the galaxy is the planar family. Therefore, in the following we restrict our study to the planar family. In the second and third rows of Fig.~\ref{fig:lyap}, we show the $xz$ and $yz$ projections of Lyapunov orbits. It can be clearly seen from the $yz$ projection that the orbit acquires some out of plane 
curvature when the parameter $\varepsilon$ increases. Moreover, the $z$ 
components of the orbit decrease with $\varepsilon$ (having in mind that we 
are showing the Lyapunov orbits about L$_2$, the opposite happens for the 
ones about L$_1$). This is, the libration point and the orbit are not
strictly contained in the plane $z=0$ when $\varepsilon\neq 0$ moreover
the periodic orbits are not strictly planar.

\begin{figure}
 
   \hspace{-0.3cm} \includegraphics[width=0.5\textwidth,angle=0]{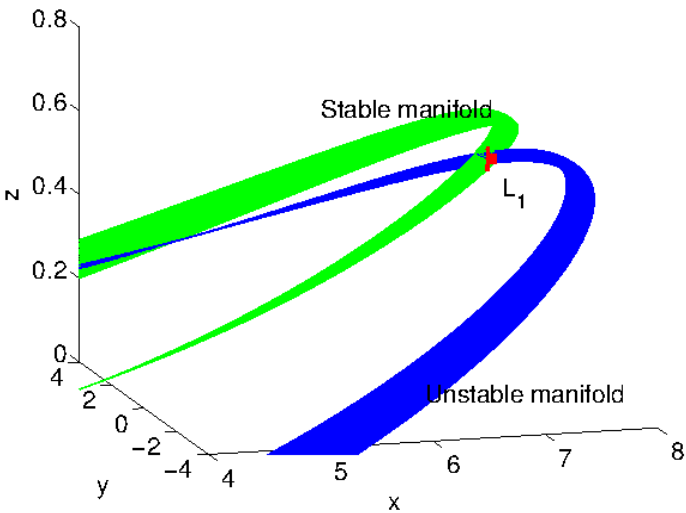}
    \caption{Stable and unstable invariant manifold associated to the periodic orbit around L$_1$ with $\varepsilon = 0.2$ and $GM_b=0.3$.}
  \label{fig:est_inest}
\end{figure}

For a given Jacobi constant, two sets of asymptotic orbits emanate from the 
periodic orbit, they are known as the stable and the unstable invariant 
manifold respectively and each set has two branches (see Fig.~\ref{fig:est_inest}). 

We denote by $W_{\gamma_i}^s$ the stable 
invariant manifold associated to the periodic orbit $\gamma_i$ around the 
equilibrium point $L_i$, $i=1,2$. This stable invariant manifold is the set 
of orbits that tend to the periodic orbit asymptotically forward in time. 
On the other hand, we denote by $W_{\gamma_i}^u$ the unstable invariant 
manifold associated to the periodic orbit $\gamma_i$ around the equilibrium 
point $L_i$, $i=1,2$. The unstable invariant manifold is the set of orbits 
that departs asymptotically from the periodic orbit (i.e. orbits that tend 
to Lyapunov orbits backwards in time). Since the invariant manifolds 
extend well beyond the neighbourhood of the equilibrium points, they are 
responsible for the global structures and the transport of matter.

\begin{figure}
  \centering
    \includegraphics[width=0.5\textwidth,angle=0]{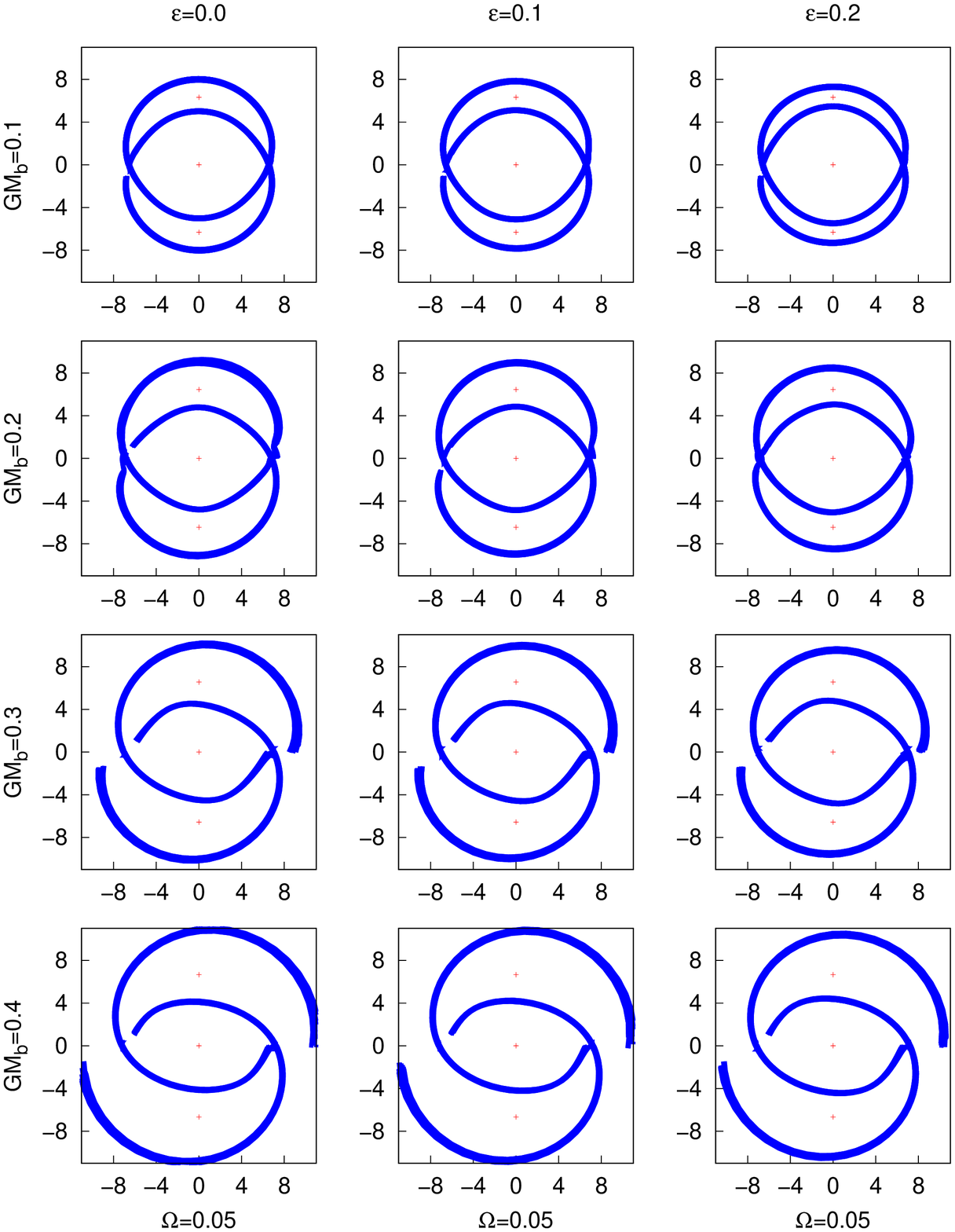}
    \caption{(x,y) projection of the unstable invariant manifolds for the precessing model with $\Omega=0.5$, $GM_b\in[0.1,0.4]$ (from top to bottom) and tilt angle $\varepsilon\in[0.,0.2]$ (from left to right). The position of the equilibrium points is marked with red crosses.}
  \label{fig:multiplot_manif_om05}
\end{figure}

\begin{figure}
  \centering
    \includegraphics[width=0.5\textwidth,angle=0]{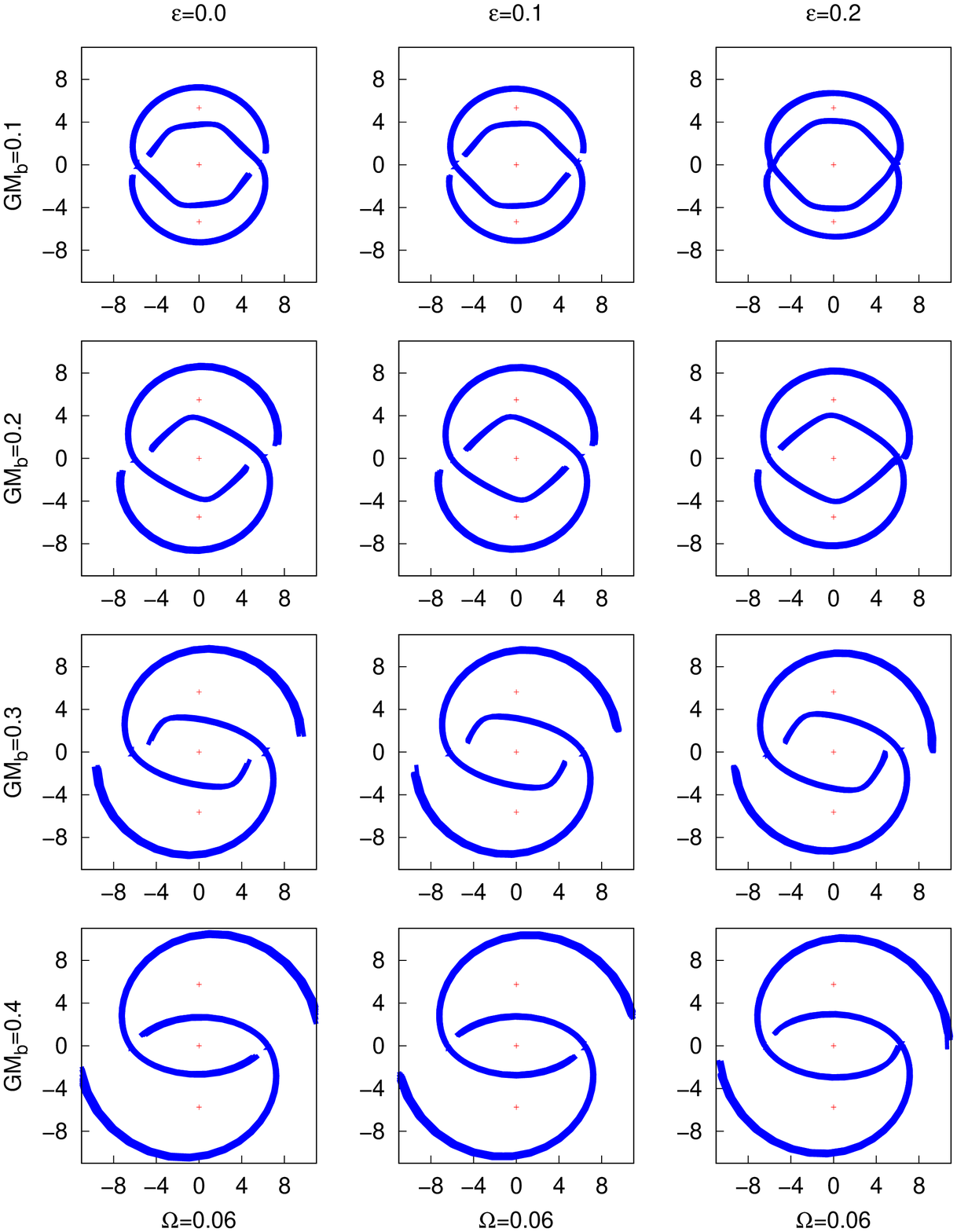}
    \caption{As in Fig.~\ref{fig:multiplot_manif_om05} for $\Omega = 0.06$.}
  \label{fig:multiplot_manif_om06}
\end{figure}

In Figs.~\ref{fig:multiplot_manif_om05} ($\Omega=0.05$) and \ref{fig:multiplot_manif_om06} ($\Omega=0.06$) we show the (x,y) projection of the invariant manifolds of Lyapunov orbits around the 
equilibrium points, $L_1$ and $L_2$, varying with the tilt angle, the angular velocity or the bar mass, for the precessing model. In both figures we have chosen the values $GM_b=0.1$ and 
$GM_d=0.9$ for the first row, $GM_b=0.2$ and $GM_d=0.8$ for the second row, 
$GM_b=0.3$ and $GM_d=0.7$ for the third row and, $GM_b=0.4$ and $GM_d=0.6$ for 
the last row. Moreover, we have set the tilt angle $\varepsilon=0$ in the 
first column, $\varepsilon=0.1$ in the second column and $\varepsilon=0.2$ 
in the last column for each figure. 


When $\Omega=0.05$ (Fig.~\ref{fig:multiplot_manif_om05}), we observe that the 
structure of invariant manifolds is preserved for different values of $\varepsilon$, but with small 
particularities. For example, the position of the invariant manifolds is not 
exactly the same in the three columns for a given value of $GM_b$. In this 
way, we can see that the structure remains but the spiral arms slowly open up. 
Moreover, when the bar mass increases, the structure moves from a morphology 
of a rR$_1$ ringed galaxy to the one of a spiral galaxy as expected \citep{Romero2}.

When we increase the pattern speed to $\Omega=0.06$, we appreciate 
(in Fig.~\ref{fig:multiplot_manif_om06}) that, although the basic structure 
is preserved, the arms are more open even for low bar masses. 
And again, the behaviour of the manifolds are the same with respect to the
variation of $\varepsilon$.

\begin{figure}
  \centering
    \includegraphics[width=0.5\textwidth,angle=0]{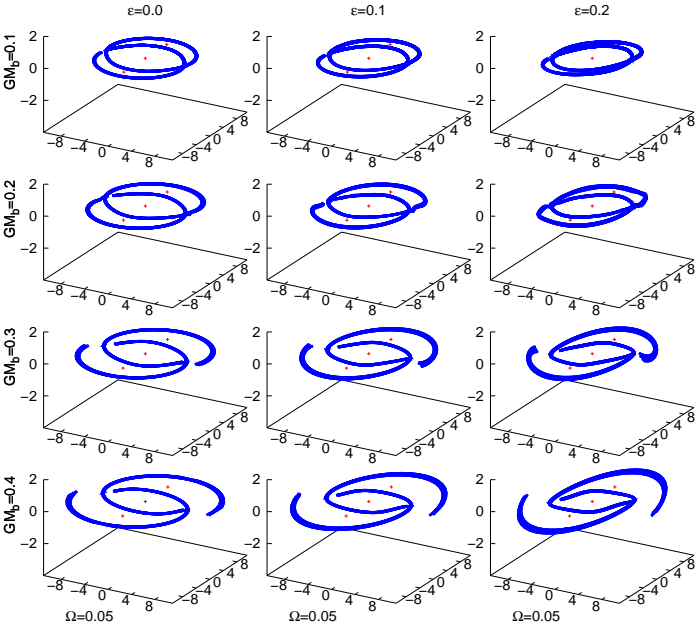}
    \caption{As in Fig.~\ref{fig:multiplot_manif_om05} in a 3D view.}
  \label{fig:multiplot_manif3D_om05}
\end{figure}

\begin{figure}
  \centering
    \includegraphics[width=0.5\textwidth,angle=0]{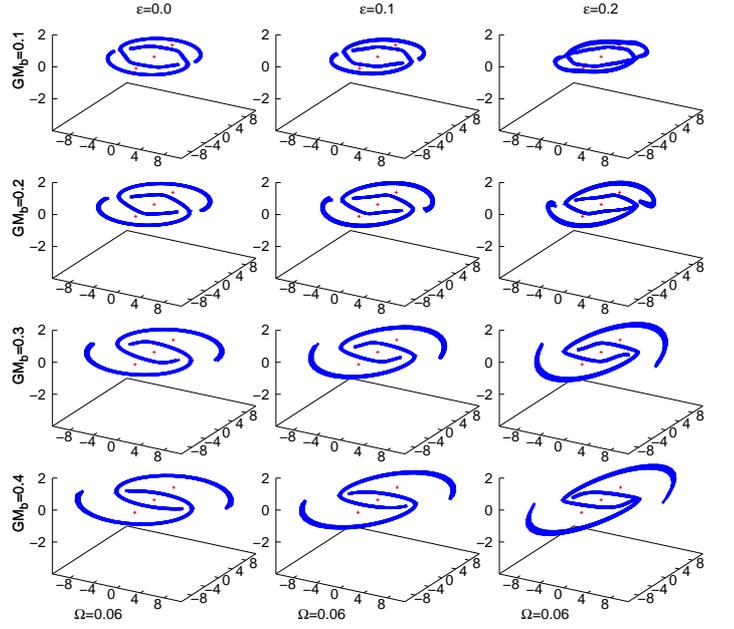}
    \caption{As in Fig.~\ref{fig:multiplot_manif_om06} in a 3D view.}
  \label{fig:multiplot_manif3D_om06}
\end{figure}

Figures~\ref{fig:multiplot_manif3D_om05} and \ref{fig:multiplot_manif3D_om06} 
show the previous panels in three dimensions, in order to better appreciate 
the variation with respect to the tilt angle of the model. Here, we clearly see 
the structures that have been discussed before, and we see how the invariant 
manifolds change in the $z$-component.

\begin{figure}
  \centering
    \includegraphics[width=0.5\textwidth,angle=0]{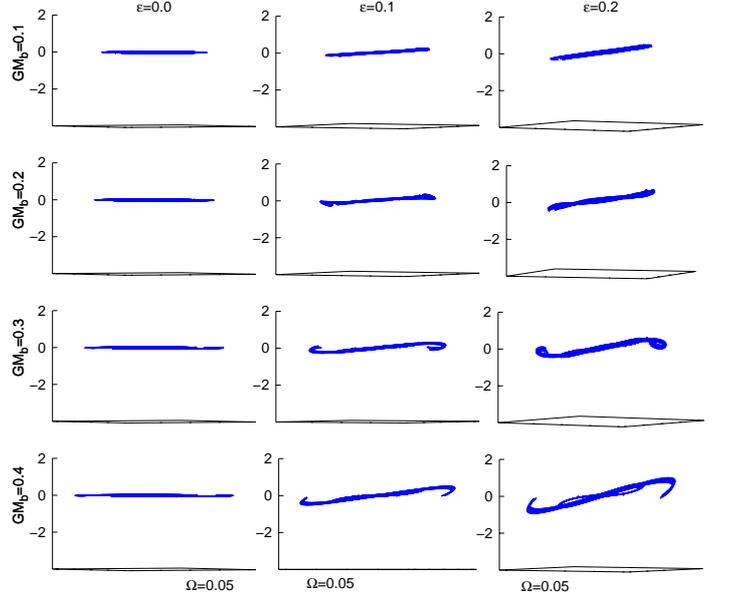}
    \caption{As in Fig.~\ref{fig:multiplot_manif_om05} but in the side-on view.}
  \label{fig:multiplot_warps_om05}
\end{figure}

\begin{figure}
  \centering
    \includegraphics[width=0.5\textwidth,angle=0]{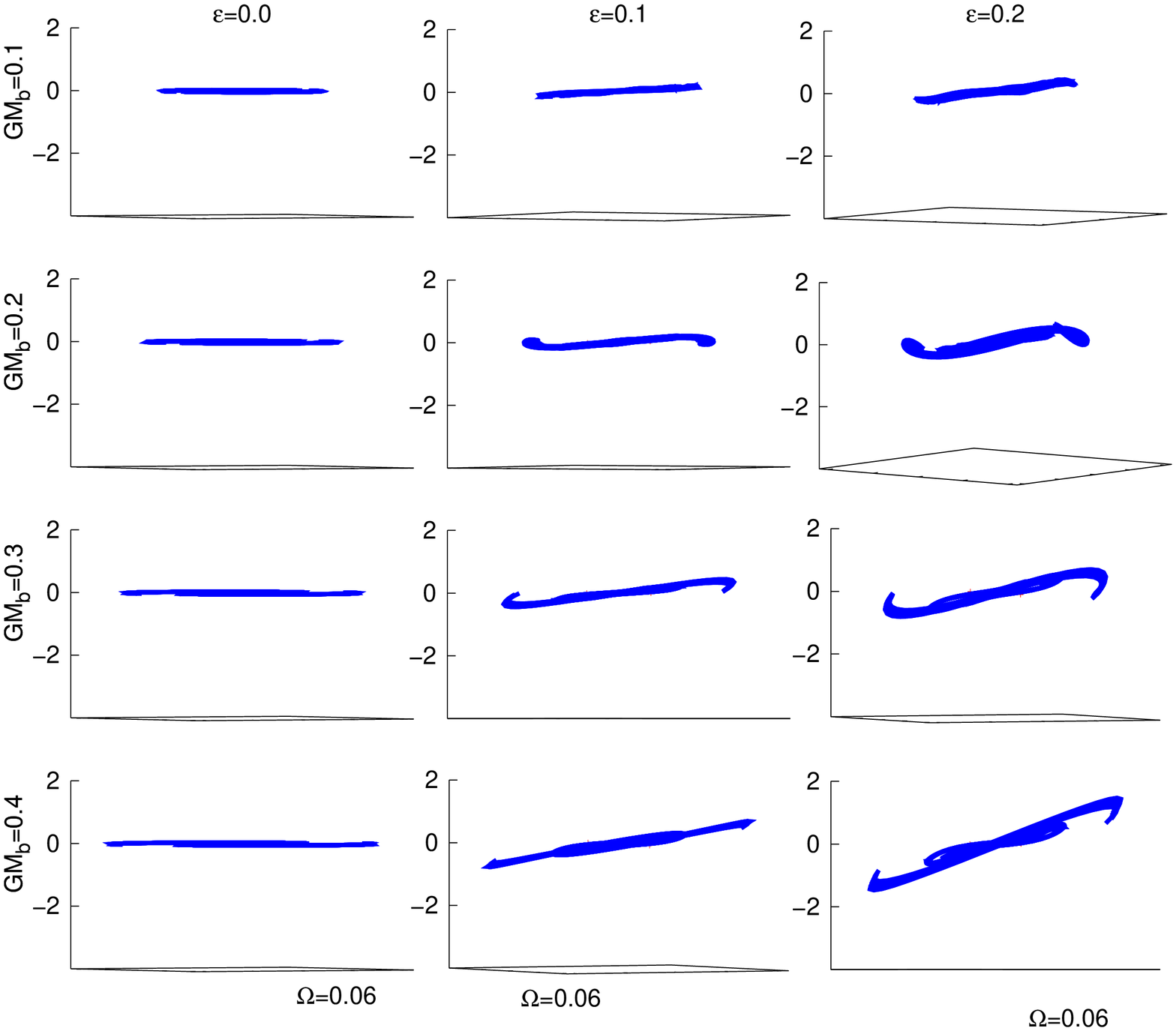}
    \caption{As in Fig.~\ref{fig:multiplot_manif_om06} but in the side-on view.}
  \label{fig:multiplot_warps_om06}
\end{figure}

In Figs.~\ref{fig:multiplot_warps_om05} and \ref{fig:multiplot_warps_om06}, 
we show a possible approach on the evidence of detected warps in galaxies. 
In a side-on view, and when the tilt angle $\varepsilon>0$, we observe 
that the outer branches of the unstable manifold are clearly warped, emulating 
the shape of some observed galaxies. For 
example, when we take a bar mass $GM_b=0.2$, all plots present a warped shape, 
and differences among them are obtained when one varies the inclination and 
pattern speed.

For $\Omega=0.05$ (Fig.~\ref{fig:multiplot_warps_om05}), we can see that no warped 
shape appears with $\varepsilon=0$ as expected. But, when $\varepsilon$ is increased, 
as well as $GM_b$, different warp shapes are present. With $GM_b=0.2$, 
invariant manifolds are just hinting the shape of warps, and when $GM_b=0.3$, the warps are clearly evidenced.

The same phenomenon occurs for the pattern speed $\Omega=0.06$ 
(Fig.~\ref{fig:multiplot_warps_om06}). But in this case the warped structure is present for a larger variety of parameter combinations. 
With $\varepsilon>0$, a warp is present with already a bar mass of $GM_b=0.2$. As $GM_b$ increases,
the S-shape of the warp becomes more evident, increasing its inclination with respect to the galactic
plane, being the most tilted case the one with $\varepsilon=0.2$ and $GM_b=0.4$. Note also that the
contribution of the inner branches of the invariant manifolds are more evident with a pattern speed faster than $\Omega=0.05$.

\begin{figure}
  \centering
    \includegraphics[width=0.5\textwidth]{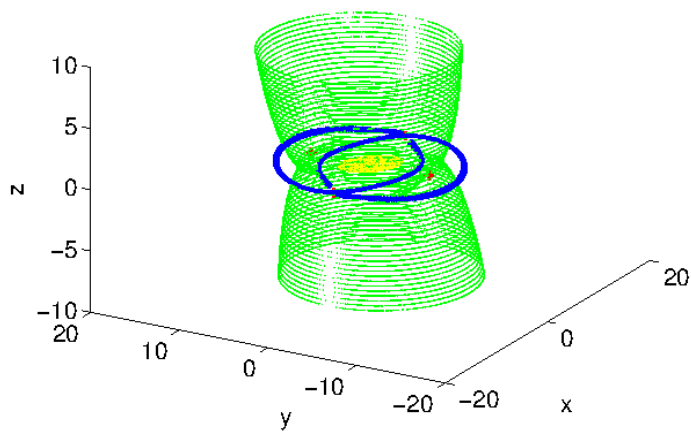}
    \caption{3D view of the unstable invariant manifolds (blue), zero velocity surface (green) and triaxial Ferrers bar (yellow) for Model S. The red crosses mark the position of the equilibrium points.}
  \label{fig:zvs}
\end{figure}

\begin{figure}
  \centering
    \includegraphics[width=0.5\textwidth]{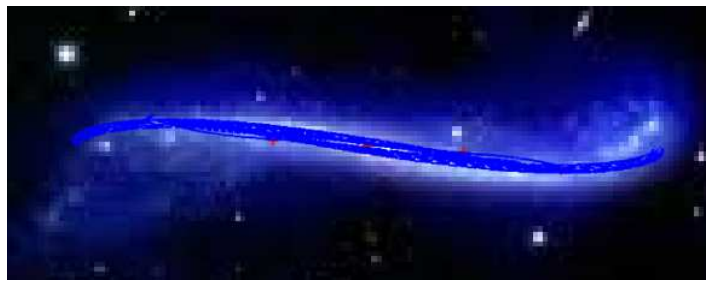}
    \caption{Warp obtained for Model S (blue) superimposed to the Integral Sign Galaxy, UGC 3697.}
  \label{fig:warp}
\end{figure}

To get a global vision, in Fig.~\ref{fig:zvs} we show the invariant manifolds for $GM_b=0.3$, $\Omega=0.05$ and $\varepsilon=0.2$ (hereafter Model S), together with the Ferrers bar and the zero velocity surface of the energy level considered. With this model, we are able to appreciate the strong resemblance with the warp of the Integral Sign Galaxy (Fig.~\ref{fig:warp}).

In summary, we conclude that the warp formation is closely related to the pattern speed 
of the bar, the bar mass and, specially, to the tilt angle of the model, as we expected. 
Also note that if we consider a symmetric bar instead of a bar with $b\neq c$, the results obtained are essentially identical. 

\subsection{Warp angles}
\label{sec:angles}
We measure the maximum amplitude of the warp as the angle between the outermost detected point and the mean position of the plane of symmetry, as defined by the internal unwarped region, as in \citet{catalog}.
The warp angle obtained in our theoretical analysis varies considerably depending on the pattern speed, bar and disc masses and, above all, the tilt angle $\varepsilon$. As we can observe in Table ~\ref{tab:ang}, when the pattern speed increases, the warp angle increases too. The same phenomenon occurs when we fix a pattern speed and we increase the bar mass, but, the biggest increment takes place when the tilt angle $\varepsilon$ grows. In this context, there are no warps when $\varepsilon=0$ as expected, but the highest warp angle, $\theta = 9.3^{\circ}$, is reached when we use the maximum values considered for all the variables. For example, for a pattern speed of $\Omega = 0.05$, the maximum angle obtained is $\theta = 7.7^{\circ}$, by setting the parameters $\varepsilon = 0.2$, $GM_b =0.4$, $GM_d = 0.6$. Whereas, if we take a pattern speed of $\Omega = 0.06$, and keep the previous values for the remaining parameters, we obtain a warp angle of $\theta = 9.3^{\circ}$. 
 
The catalogue of warps in the Southern hemisphere \citep{catalog} shows that most warps have angles less than 11$^{\circ}$, which is very close to the maximum warp angle of $\theta = 9.3^{\circ}$ we have obtained with our theoretical model. Let us notice that the tilt angle $\varepsilon$ has to be small, since otherwise the system would lose its consistency, in the sense that if the tilt angle were bigger, the model would be unstable, and it would lead to chaotic dynamics. The model is stable up to tilt angles slightly above $\varepsilon = 0.25$ rad, which could produce warp angles close to 11$^{\circ}$.

\begin{table}
\centering
\caption{Warp angles (in degrees) obtained in the precessing model.}
\begin{tabular}{||c | c | c | c||}
\hline
\hline
$\varepsilon$ & $\Omega$ & GM$_b$ & $\theta$ ($^{\circ}$) \\
\hline
0.1 & 0.05 & 0.1 - 0.4 & 1.8 - 3.9\\
\hline
0.1 & 0.06 & 0.1 - 0.4 & 1.8 - 4.8\\
\hline
0.2 & 0.05 & 0.1 - 0.4 & 3.8 - 7.7\\
\hline
0.2 & 0.06 & 0.1 - 0.4 & 3.7 - 9.3\\
\hline
\end{tabular}
\label{tab:ang}
\end{table}

\section{Test particle simulation} \label{sect:section5}
The advantage of test particle simulations is that the stars are evolved 
using a known galactic potential and they have inherited the information 
on both density and kinematics, that is, the stars are in statistical 
equilibrium with the potential imposed after a certain integration time. 
They are used as generators of mock catalogues \citep{Romero4} or to obtain 
information of the potential imposed by studying certain aspects of the 
simulation, for example the moving groups in the Solar Neighbourhood 
\citep[e.g.][]{Dehnen,fux01,gar10,min10,ant11}.

The purpose of using test particle simulations here is twofold: first, to show that the particles are trapped in the manifolds when integrating in the precessing model and present a warped shape. Second, to show that not only the manifolds warp, but also the orbits in the disc present a warped shape. The manifolds, being the backbone of the spiral arms, contribute to it. Therefore, we generate a set of $10^6$ particles
using the Hernquist method \citep{her93}. The density follows the same
Miyamoto-Nagai disc \citep[see Appendix of][]{Romero4} as in the analytical
computations. We give the particles the initial velocity for a circular orbit
with zero dispersion. 
The bar pattern speed is set to $\Omega=0{.}05$ [u$_t$]$^{-1}$, i.e. $1$ bar rotation takes $125$ Myr. The bar is introduced adiabatically in $t_1=16$ bar rotations, using the same time function as in 
\citet{Dehnen} in the precessing model:  
 \begin{equation}
  A_b=A_f\left(\frac{3}{16}\xi^5-\frac{5}{8}\xi^3+\frac{15}{16}\xi+\frac{1}{2}\right),\, \xi\equiv 2\frac{t}{t_1}-1, \,t \in (0,t_1),
 \label{eqn:amplit}
 \end{equation}
and $A_b=0$ if $t\leq0$. $A_b$ grows with time in the interval 
$t \in (0,t_1)$, and assumes its maximal amplitude when $t\ge t_1$, 
in which $A_b=A_f$, that is, it assumes the total bar amplitude. 
Since Eq.~(\ref{eqn:amplit}) is continuous and 
derivable, a smooth transition from non-barred to a barred galaxy is guaranteed.

In order to keep the total mass of the system constant when we introduce the bar adiabatically, we transfer mass from the disc to the bar progressively, so that
\begin{equation}
 \phi_T = (1-f(t)f_0)\phi_d+f(t)\phi_b,
\end{equation}
where $\phi_T$ is the total potential of the system, $\phi_d$, $\phi_b$ the potentials of the disc and bar, respectively, and the time function $f(t)$ is the same polynomial of time $t$ as in Eq.~(\ref{eqn:amplit}). The parameter $f_0$ takes the value of the final bar mass, $f_0=GM_b=0.3$, so that when the integration time reaches the maximum amplitude of the bar, $t=t_1$, the bar mass is $GM_b=0.3$ and the mass disc $GM_d=0.7$. Thus, we consider Model S once the final configuration is reached. 

\begin{figure}
 \centering
 \includegraphics[width=0.4\textwidth]{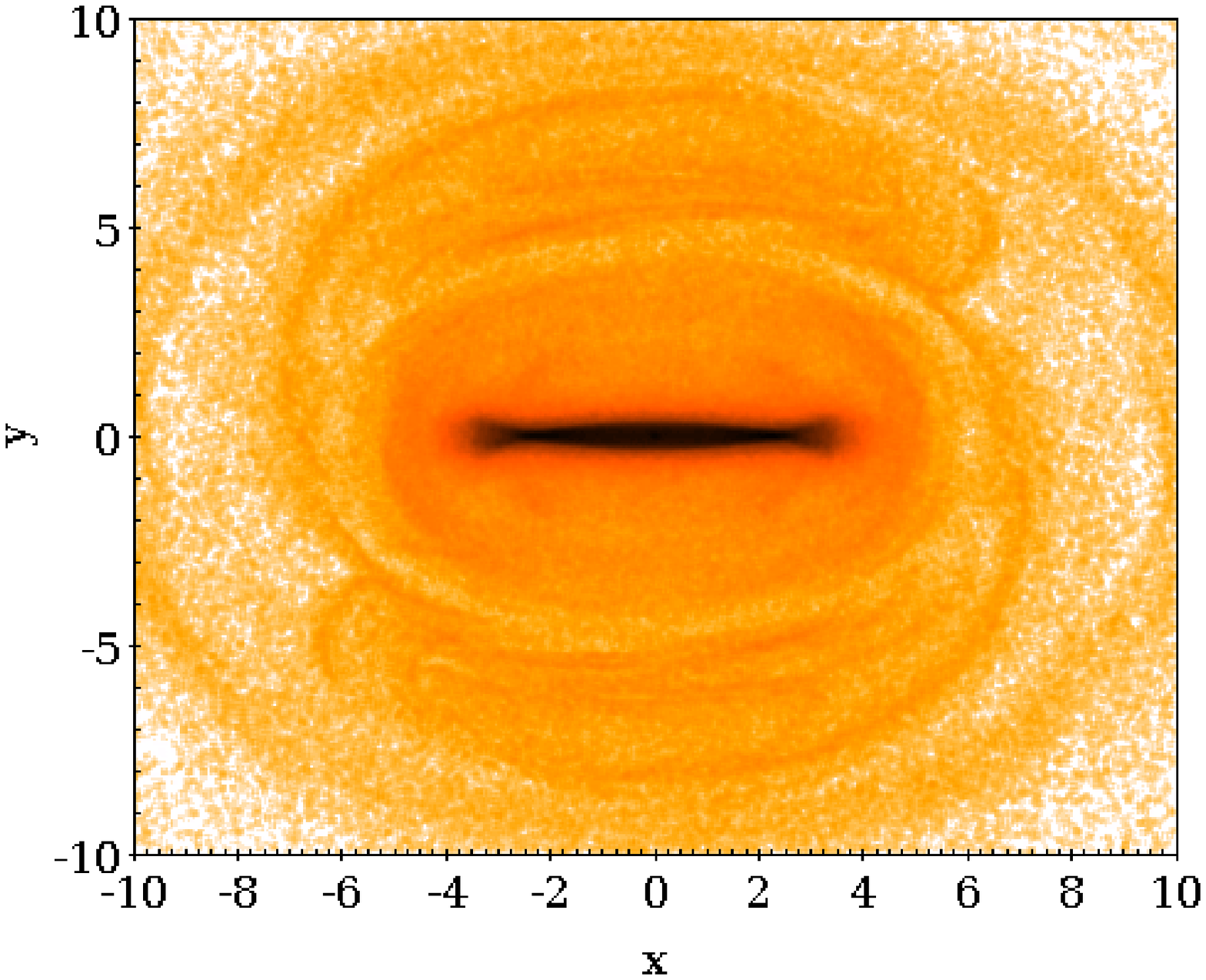}
  \includegraphics[width=0.4\textwidth]{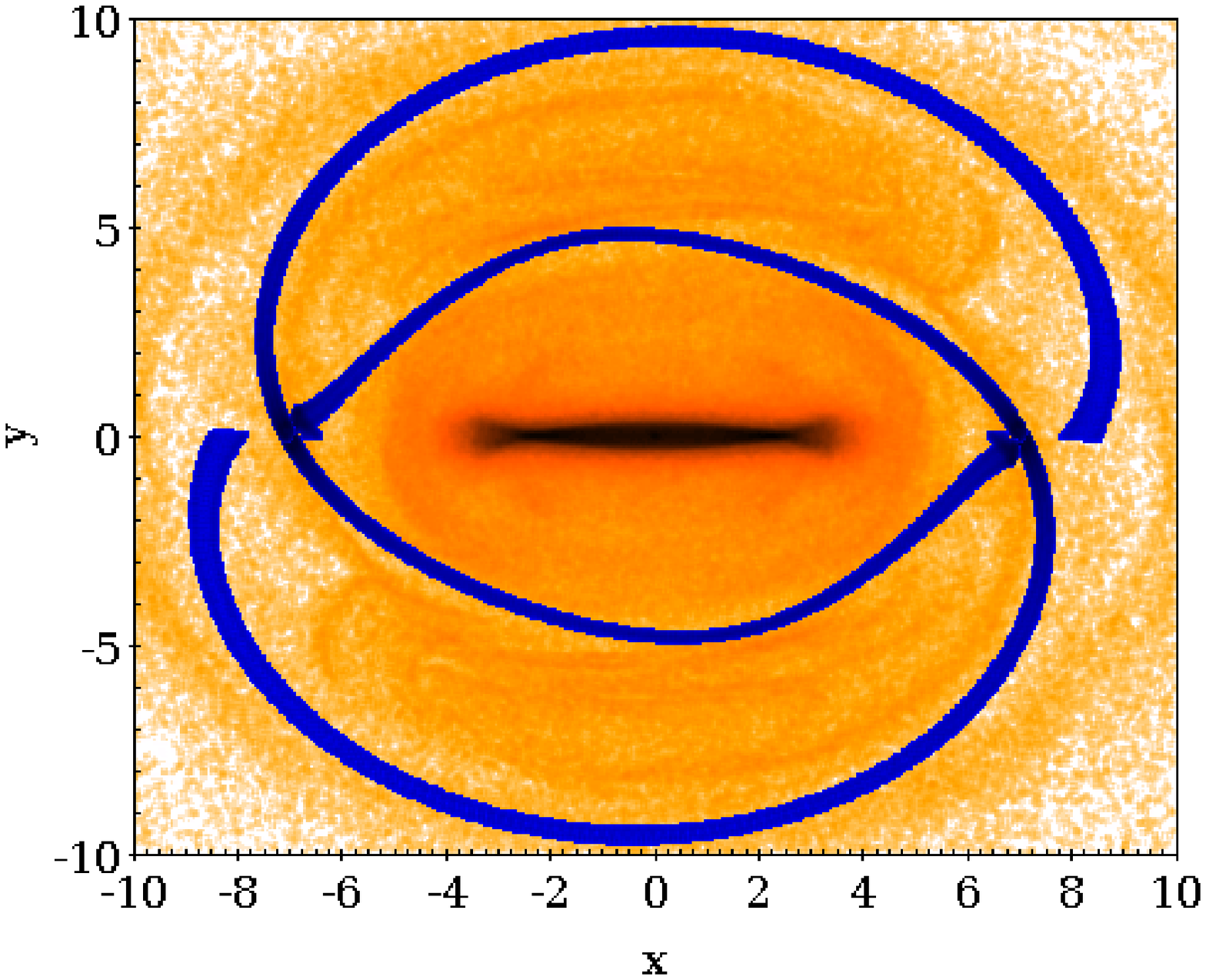}
  \caption{ Top: Surface density of the $xy$-projection of the test particle simulation for Model S. Bottom: Overlap of  
  the invariant manifolds for the same parameters (in blue) with Jacobi constant $C_J=-0.19366$.}
 \label{fig:disp0xy}
\end{figure}

The particles in the $xy$ plane adopt the shape seen in Sect.~\ref{sect:section4}. The top panel of Fig.~\ref{fig:disp0xy} presents the configuration of the particles, which acquire characteristic features. We observe how some particles are concentrated in the 
$L_4$ ($L_5$) region, whose stable family of periodic orbits prevents
these particles from exiting the region. Even though the $L_4$ and $L_5$
Lagrangian points remain always in the galactic plane, the orbits around them
are non-planar and they slightly contribute to the warped shape. 
Also, the particles in the outer parts of the
zero velocity curves adopt the shape of the invariant manifolds as we expected.
This is shown in the bottom panel of Fig.~\ref{fig:disp0xy}, where, for the
selected model, we overlap the invariant manifolds of the unstable orbits
around $L_1$ and $L_2$ with a Jacobi integral $C_J$ close to that of the
equilibrium point $L_1$ ($C_J=-0.19366$ and $C_{J,L_1}=-0.19368$).


\begin{figure}
 \centering
 \includegraphics[width=0.5\textwidth]{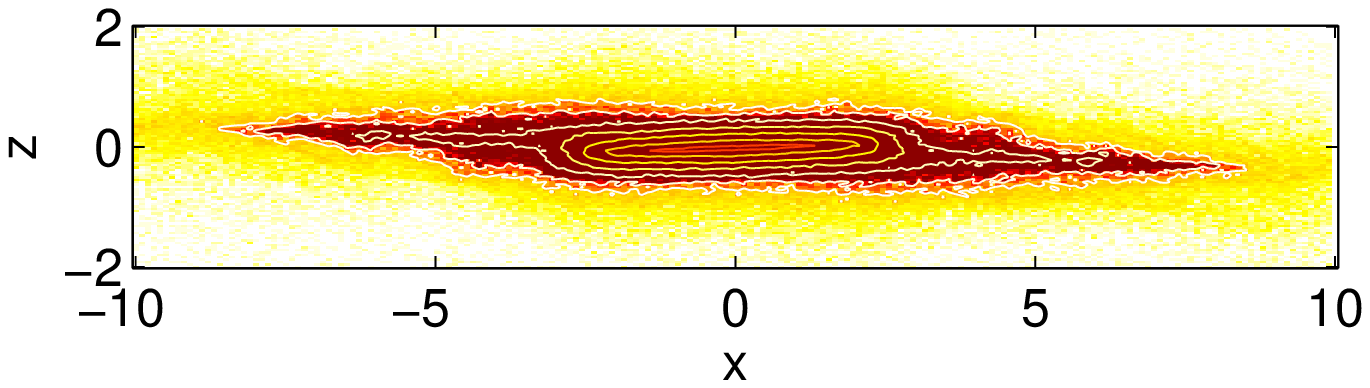}
  \includegraphics[width=0.5\textwidth]{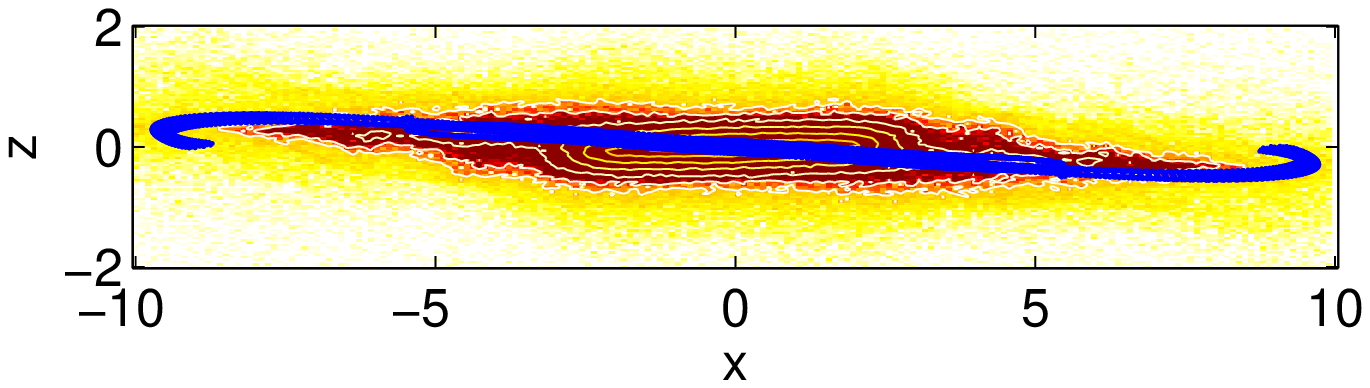}
  \caption{ Top: Surface density and contours of the $xz$-projection of the test particle simulation for Model S. Bottom: Overlap of  
  the invariant manifolds for the same parameters (in blue) with Jacobi constant $C_J=-0.19366$.}
 \label{fig:disp0}
\end{figure}

However, not only the particles trapped in the manifolds contribute to warping the disc. In Fig.~\ref{fig:disp0} we show the surface density and its contour levels in the $xz$-projection of the test particle simulation (top panel) and we overlap the invariant manifolds of the unstable orbits around $L_1$ and $L_2$ (bottom panel). Note, first, how the precessing model tilts the bar and disc and it evidences a warped shape towards the outer parts. Note also the overdensity due to the superposition of the bar and the particles trapped by the outer branches of the invariant manifolds. This overdensity is not only due to the invariant manifolds. In other models it can be due to other families of periodic orbits that are trapped in the vertical resonances forming a thick spiral (see e.g. \citet{Kalnajs, pat96}).

If we compare the density contour with others found in the literature, as for example the one shown in Fig. 3 of \citet{DebSell} obtained from a N-body simulation, we observe that the tilting of our model is evidently acquiring a similar shape to that in the mentioned figure.

We can also compare our results to that of observations. As previously mentioned, the invariant manifolds of this model match the profile shown by the Integral Sign Galaxy (Fig.~\ref{fig:warp}). In this case, the maximum angle of the warp is 
$\theta=6.7^\circ$. This value is in agreement with warp angles observed in 
external galaxies \citep{catalog}, as discussed in the previous Sect.~\ref{sec:angles}. Using the test particle simulation, we can see that, indeed, the particles integrated in the precessing model get warped in a similar way to that of the Integral Sign Galaxy.


\section{ Discussion} \label{sect:section6}

Although warps are a common feature in galaxies, there is still no agreement on
how the warp is formed. In this paper we continue an idea which started 
in~\citet{Romero1}. It is based on the fact that spirals and rings in barred
galaxies can be driven by the invariant manifolds associated to the unstable
Lyapunov periodic orbits around the unstable equilibrium points in the rotating
bar potential. Here we investigate the effect of tilting the model with respect
to the $XY$ plane, which is equivalent to a small misalignment between
the angular momentum and the angular velocity of the system. Since invariant
manifolds behave like tubes transporting matter, we have been able to observe
that due to the small misalignment these manifolds reproduce warped shapes as
observed in warped galaxies. When we study the motion of test particles under 
the precessing model, we see that even though the main contribution comes from the 
invariant manifolds, all orbits warp. In addition, we have shown the consistency of the
model despite its tilting thanks to the periodic orbits inside the bar (around
the central equilibrium point), which continue being responsible for the bar
structure and constitute its backbone.

To make sure that our principal results are not model-dependent, we prove our
theory with a model composed by a bar with revolution symmetry and another one
with just axial symmetry. In both models, the periodic orbits around the
central point $L_3$ contribute to the backbone of the bar, the family of
periodic orbits being qualitatively the same in both cases, and although the shape of
the stability indexes varies, the range of energies in which the orbits are
stable is the same. But the main point is that the invariant manifolds in both
models are very similar and acquire the same warped form, in the sense that if
we take equal values for the free parameters (bar mass, bar disc, tilt angle
and pattern speed) we obtain equivalent warped shapes.

The addition of a spherical dark matter halo to the galactic model has been studied
in detail in~\citep{patsan}, where it leads to the same results
as in this work. The position of the equilibrium points varies with the
presence or absence of the halo, and with its mass. But the behaviour of these
points does not change: the two equilibrium points at the ends of the bar
continue being unstable, whereas the rest remain stable. As for the invariant
manifolds, the halo affects to a greater extent the inner than the outer
branches of the invariant manifolds. An increase of the mass of the halo makes
the inner branches join and the outer ones to become more open, while if its 
mass decreases the inner branches open up forming a ring and the outer branches
slowly close. The halo also influences the formation of warps, favouring 
larger warp amplitudes, but within observational ranges.

Comparing with observations, we can confirm that the warp angles obtained with
this precessing model closely approximate observed warps. We observe that the
tilt angle $\varepsilon$, which is the angle between the angular momentum and
the angular velocity is also responsible for the warp shape, though, the warped
shape also depends on the pattern speed and bar mass, albeit to a second order.
We show that if the bar mass grows or the pattern speed is faster, the warp
angle increases. In the precessing model, the warp begins close to the corotation radius, where the Lagrangian points are located, and it is related to the warped invariant manifolds. In external galaxies, it is believed that the galaxies are flat within $R_{25}$ and warps become detectable within the Holmberg radius, $R_{Ho}=R_{26.5}$ \citep{Briggs}, which are the radii of the isophote of an elliptical galaxy corresponding to a surface brightness of $25$ and $26.5$, respectively, blue magnitudes per square arcsecond. It is difficult to test whether the Holmberg radius is close to the corotation radius because there are few edge-on warped galaxies classified as spiral barred galaxies. In some galaxies we can obtain the ratio $R_{bar}/R_{25}$, which is $0.1$ for NGC3344 \citep{Ver00}, $0.19$ for M33 \citep{Elm92,Her09} and $0.37$ for NGC~5560 \citep{Bai11}. These values indicate that the warp begins far from the end of the bar, however, the relation between $R_{bar}$ and the corotation radius depends on the bar pattern speed. If it is a slow rotator, the corotation radius moves farther out and it can be close to $R_{25}$, or at least within the area the spiral arms cover.

The warps generated in N-body simulations have a different origin from the one proposed here. They range from a reorientation of the outer halo caused by cosmic infall \citep[e.g.][]{Jiang,infall}; a flyby scenario, that is caused by an impulsive encounter between two galaxies \citep{kim14}; bending instabilities \citep{bend_warp}; an external tidal torque causing a tumbling misaligned halo with the disk \citep{dub09}; such misalignments can be also between the inner disc and the hot gaseous halo \citep{ros10} or between the angular momenta of the disk and halo \citep{DebSell} or between the principal axes of the triaxial halo \citep{hu15}. The warps obtained in the works mentioned above have very similar characteristics and the warp angles obtained are comparable to the ones from observations, although in some cases, the warp angle remains in the lower ranges \citep[e.g.][]{bend_warp,kim14}. Not all simulations are cold enough to form non-axisymmetric structures like bar and spiral arms, such as in the cosmological simulation from \citet{ros10}, however, in other cases, a spiral structure is present in the outer parts of the disc as well as the warp \citep{DebSell,dub09,kim14,hu15}. In the recent work of \citet{hu15}, the authors claim that the warp interferes very little with the spiral structures. This fact also happens in our precessing model, where we showed that the $XY$ projection remains the same as if no misalignment between the angular momentum and angular velocity exists. \citet{dub09} show that the disk behaves as a rigid body and that stars in the outer parts of the disc, with weaker self-gravity are the ones that precess differentially and form the warp.

Finally, let us point out that this work is a first approximation, establishing
which parameters in our model are related to the formation of warps. A more
detailed classification of the warps obtained will be the subject of future
investigation.

\section*{Acknowledgements}

This work has been partially supported by the MINECO (Spanish Ministry of 
Economy) - FEDER grants AYA2012-39551-C02-01 and ESP2013-48318-C2-1-R, 
MTM2012-31714 and the Catalan Grant 2014SGR504. PSM has been supported by 
the Catalan PhD grants FI-AGAUR and FPU-UPC. The test particle simulations have run in the clusters EIXAM (at MA1-UPC) and Sol (at DAM-UB). We thank the anonymous referee for constructive comments that helped improving the manuscript.

\end{document}